\documentstyle[11pt,aaspp4,flushrt]{article}
\begin{document}

\title{Simulation of Stellar Objects in SDSS Color Space}

\author{Xiaohui Fan}
\affil{Princeton University Observatory\\
Princeton, NJ, 08544 \\
Email : fan@astro.princeton.edu}

\begin{abstract}
We present a simulation of the spatial, luminosity and spectral
distributions of four types of stellar objects. 
We simulate:
(1) Galactic stars, based on a Galactic structure model, a stellar population
synthesis model, stellar isochrones, and stellar spectral libraries;
(2) white dwarfs, based on model atmospheres, the observed luminosity function, mass distribution,
and Galactic distribution of white dwarfs;
(3) quasars, based on their observed luminosity function and its evolution,
and models of emission and absorption spectra of quasars;
and (4) compact emission line galaxies, based on the observed distribution of
their spectral properties and sizes. 
The results are  presented in the color system of the Sloan Digital
Sky Survey (SDSS), with realistic photometric error
and Galactic extinction. 
The simulated colors of stars and quasars are compared with observations in 
the SDSS system and show good agreement.
The stellar simulation can be used as a tool to analyze star counts
and constrain models of Galactic structure, as well as 
to identify stars with unusual colors.
The simulation can also be used to establish the quasar 
target selection algorithm for the SDSS.

\end{abstract}

\section{Introduction}

The Sloan Digital Sky Survey (SDSS, see e.g. Gunn \& Knapp 1993,
Gunn \& Weinberg 1995)
will obtain  deep CCD images of about $\pi$ steradians (10000 deg$^2$) of sky
centered on the North Galactic Cap in five bands, as well as
spectra of approximately 1 million galaxies, 100,000 quasars
and numerous stars selected from the  photometric observations.
It will detect about $10^{8}$ stellar objects and $10^{8}$ galaxies
down to its limiting magnitude ($r' \sim 23.1$). 

The five photometric bands
($u', g', r', i'$ and $z'$) of the SDSS 
cover the entire optical range from the UV atmospheric 
cutoff at about 3000 \AA\ to the red silicon sensitivity cutoff at about 10000 \AA\ (Fukugita {\em et al.} 1996, F96 hereafter).
The SDSS photometric system is based on the AB$_{95}$ system defined 
in F96.
The zeropoints of each passband is defined by the observations of
Oke-Gunn (1983) spectrophotometric standard stars.
Details of the photometric system is given in F96 and in Appendix A.
The sensitivity curves of the five bands (including the contributions 
from the CCD, filter,  telescope optics and average atmosphere
transmission, see also Appendix A) are shown in Figure 1.
The spectra of a typical $z=2.8$ quasar and an F5V star are also plotted
on Figure 1. The striking similarity of the overall shapes
of these two spectra reveals  the well-known difficulty of separating 
quasars at $z \sim 2.5 - 3.0$ from normal stars using only broad-band colors (\S 6).

The scientific goals of the SDSS require a detailed simulation of the distribution of 
objects in color space, as a function of position on the sky.
In this paper, we present simulations of four kinds of stellar objects 
that SDSS will observe: normal Galactic stars, white dwarfs, quasars,
and compact emission line galaxies (CELGs) the software will classify as point sources.
The work is motivated by two important scientific projects of the SDSS:

1. Quasar Selection. SDSS will select quasar candidates according to both
their morphology (point source) and colors.
It will generate the biggest and most homogeneous quasar sample to date, 
covering redshift $z$ from 0 to $>5$.
Quasars will be selected as the outliers  
from the dense stellar locus in color space
by the target selection pipeline (Newberg \& Yanny 1997).  
However, objects such as white dwarfs, CELGs and very blue or red stars
also have colors different from normal stars, and
are located outside of the stellar locus.
Selecting them as quasar candidates will lower the success rate of
the quasar spectroscopic identifications.
The quasar selection criteria need  
(a) to produce  a high success rate for quasars,
i.e., to exclude possible non-quasar contaminants as much as possible (\S6.3);
(b) to select quasars as uniformly as possible, as a function of
both sky position and redshift; 
and (c) to select the quasar sample as complete as possible.
The selection criteria will be determined using a limited amount of data
during the commissioning period of the survey. We therefore need detailed
modeling of the distribution of stellar objects in SDSS color space,
especially  in the following senses: (a) modeling of the quasar colors as a
function of redshift, (b) modeling of the distributions of possible
contaminants of quasar candidates: white dwarfs, CELGs
and stars, both in color space and in different
directions on the sky; and (c) modeling of the photometric errors of the SDSS,
this is especially important for selecting faint quasars and high redshift quasars.

2. Galactic Structure. Half of the sources that the SDSS will detect will be
stars in our Galaxy.
The SDSS covers the whole Northern Galactic Cap, roughly speaking,
with Galactic latitude $b > 30^{\circ}$.
Compared to previous data used for star count studies, 
it covers a much larger part of the sky, with 5 well-calibrated bands
(see Appendix A),
and high photometric accuracy. 
SDSS is especially suitable for constraining Galaxy models, in particular
the properties of the halo and thick disk; finding distinct groups or 
low surface brightness dwarf galaxies (like the Sagittarius dwarf) in the halo;
and determining the distributions of interesting classes of stars like
white dwarfs and  carbon stars. 
All these studies require detailed modeling of the colors and spatial
distributions of stars. 

The overall structure of the simulation presented in 
this paper  is illustrated in Figure 2.
It has five independent parts: one general part to calculate and apply
Galactic extinction and 
random photometric errors in the SDSS system;
and four special parts to simulate  each of the four
kind of stellar objects we consider in this paper. 

A point source model (SKY) that predicts number counts of Galactic and
extragalactic sources is developed by Cohen (1993,
see also Wainscoat {\em et al.} 1992, Cohen 1994).
SKY predicts number counts over a wide range of wavelengths from
1400 \AA\ to 35 microns based on synthetic spectral library of
87 categories of sources.
While the simulation presented in this paper has been
fine-tuned to the SDSS observations as well as to the studies of
Galactic structure and quasar selection from SDSS data,
SKY will further provide valuable information on the classes of objects
we do not attempt to simulation here (\S 2.4).

The paper is organized as follows:
we discuss the simulations of stars, white dwarfs,
quasars and CELGs in \S 2 -- 5, respectively,  
following the structure shown in Figure 2.
In each of these sections, we first discuss the spectral models of
those objects, then discuss the population models 
(the spatial and luminosity distributions), and finally 
show the simulation results and compare them with test observations
from the literature.
In \S 6, we first describe the distributions of stellar objects in the SDSS color
space in general, based on the simulations presented in the
previous sections;
then we discuss the application of 
the simulation to quasar selection.
We summarize the major conclusions of the paper in \S 7.
In the appendices, we provide more technical details of the simulation
tools, including the realization of photometric errors
(Appendix A), and simulation of 
quasar absorption systems (Appendix B).

\section{Simulation of Galactic Stars}

Photometric surveys have been used as an important tool
to study Galactic structure and the evolutionary history
of our Galaxy (for reviews, see Bahcall 1986 and Majewski 1993).
Many computer models have been developed to interpret those data. 
Among them are  the classic star count models by Bahcall \& Soneira (1980, 1984,
BS hereafter),
as well as Gilmore (1984) and Reid \& Majewski (1993). 
These simulations are based on a Galactic model (disk+halo, sometimes with thick disk),
and the observed luminosity function and color-magnitude diagram of each population. 
Works by Ng and collaborators based their simulated catalog on
population synthesis models assuming the star formation history
of our Galaxy (Ng 1994, Ng {\em et al.} 1997). Haywood {\em et al} (1997a, b, 
see also Robin \& Cr\'{e}z\'{e} 1986, Haywood 1994) based their models
on the synthesis of both the evolution of stellar population and
the dynamical evolution of vertical structure of the disk.

Our Galactic simulation is based on :
\begin{enumerate}
\item the hybrid model stellar atmosphere libraries
compiled by Lejeune {\em et al.} (1997a, b). We use them to calculate
the SDSS magnitudes of stars given the effective temperature $ T_{eff}$,
metallicity [Fe/H], and surface gravity $\log g$; 
\item a simple stellar population synthesis model based 
on the Padova stellar evolutionary tracks and isochrones (Bertelli {\em et al.} 1994).
We use it to generate the simulated catalog of a given stellar population;
\item  a BS type Galactic structure model 
(with some modifications), including a Galactic disk with scale height
that grows in time, and a flattened Galactic halo.  We use it to generate the
properties of stellar populations as a function of Galactic position.
\end{enumerate}

\subsection{Stellar Libraries}

We use the grids of theoretical stellar atmospheres compiled by 
Lejeune {\em et al.} (1997a, b). 
The grids are based on three different sets of models covering
different parts of parameter  space: 
(1) Kurucz (1995) models in the temperature range of 3500 -- 50000 K;
(2) Allard \& Hauschildt (1995) M-dwarf models in the temperature
range of 2000 -- 3500 K and $\log g >3$; and (3) Fluks {\em et al.} (1994)
synthetic spectra of M giants in the temperature range of 2500 -- 3500 K
and $\log g <$ 3. The whole library covers the following parameter range:
$T_{eff}$ : 2000 -- 50000 K; log $g$ : --1 -- +5.5; [Fe/H] : --3.5 -- +1.0.
We then calculate the SDSS magnitudes of those grid  spectra
following Appendix A.
We interpolate between the grids linearly in $\log T_{eff}$, log $g$
and [Fe/H].
For a compilation of SDSS magnitudes based on Kurucz (1995) models, 
see Lenz {\em et al.} (1998).

Atmospheres of late type stars are difficult to model.
There is a noticeable discontinuity in the model stellar colors 
at  $T_{eff} \sim$ 3500 K, the transition region from
Kurucz models to those of  Allard \& Hauschildt (see, e.g., Figure 4c).
Lejeune {\em et al.} (1997a,b) further applied an empirical 
correction to the original atmosphere grids, so that the synthetic
UBVRI colors matched the empirical color-temperature calibrations
from observed cool stars.
This smooths the transition between models.
However, the smoothed models produce a redder $g'-r'$ color for very cool stars
than the observations of Newberg {\em et al.} (1998),
possibly due to the differences between the SDSS bands and
the UBVRI bands where those empirical corrections were derived.
We therefore retain the original Kurucz and Allard \& Hauschildt 
grids, and leave the problem of calibrating the cool star color-temperature
relation to further SDSS observations.

We use the Padova library of isochrones (\S 5.2 of Leitherer {\em et al.} 1996).
The isochrones are calculated from the stellar evolution models
of the Padova group (for $\rm M > 0.6 M_{\odot}$, Bertelli {\em et al.}
1994) with the aid of the VandenBerg (1985) models for $\rm 0.15  M_{\odot} 
< M < 0.6 M_{\odot}$.  
The grids cover the metallicities of $Z$=0.0004, 0.004, 0.008, 0.02,
0.05 and 0.1.
They correspond to [Fe/H] = --1.7 -- +0.7 assuming $Z_{\odot} = 0.02$.
Each evolutionary sequence is followed from the Zero Age Main Sequence (ZAMS)
to the beginning of the thermally pulsing region 
of the AGB phase for low and intermediate mass stars. 
We interpolate between the grid points 
linearly in Z and log(Age) of the isochrones.  
For star with a given age, in the mass direction, we interpolate linearly in $\rm M_{0}/M_{0,TO}$,
where $\rm M_{0}$ is the ZAMS
mass of the star, and $\rm M_{0,TO}$ is the ZAMS mass of
the main sequence turnoff for that age.

\subsection{Single Population Synthesis}

In principle, one can simulate a stellar population from one of the
two approaches. BS  (see also
Reid \& Majewski 1993) simulate  stars based on  observed
luminosity functions (LFs) and color-magnitude diagrams (CMDs).
This approach is sufficient for star count studies, especially those
based on two or three colors. The SDSS observations, however,
cover five optical bands, with $u'$ color sensitive to the changes
in metallicity and luminosity class (or $\log g$, see also
Lenz {\em et al.} 1998). 
The LF doesn't contain this information. 

Therefore, we adopt an approach based on evolutionary stellar population
synthesis, similar to the models described in Haywood {\em et al.} (1994a,b)
and Ng {\em et al.} (1997), using the observed properties of
the initial mass function (IMF), star formation rate (SFR) and the 
age-metallicity relation (AMR).
Our simulation consists of two distinct populations: a Galactic disk
with scale height growing in age, and a Galactic halo.
A third population, a thick disk, can be added easily if needed. 
The physical parameters characterizing each population are 
summarized in Table 1.

1. Disk.

Neither the IMF nor the SFR of the disk is  well-determined 
observationally (Majewski 1993, Scalo 1998). 
We require our adopted IMF and SFR to generate a theoretical LF
that is consistent with the observed LF (as used by BS).
We adopt the IMF and SFR of Haywood {\em et al.} (1994a,b).
The age of the disk is 10 Gyr and the SFR is assumed to be constant 
in time for the past 10 Gyr .
The IMF is a broken power law with:

\begin{equation}
dN/dM \propto M^{-(1+x)}, x=\left\{ \begin{array}{ll}	
			0.7 & M < 1 M_{\odot} \\
			1.5 & 1  M_{\odot} < M < 3 M_{\odot} \\
			2.0 & M > 3 M_{\odot} 
				\end{array}
				\right.
\end{equation}

\noindent
This form is consistent with the observed IMF of Kroupa {\em et al.} (1993).
Note that for the old disk stars, the detailed slope at the high mass end
is not important.  

We adopt the relation between metallicity and age of Rana (1991):

\begin{equation}
[\rm Fe/H]_{mean} = 0.68 - 11.2/(20-\rm t), \rm  \sigma([Fe/H]) = 0.20
\end{equation}

\noindent
where  t is the age of the star in units of Gyr, 
and $\rm [Fe/H]_{mean}$ is the average
metallicity of that age, with a Gaussian scatter of 0.20.
The disk has an average [Fe/H] of --0.44 when it was formed (10 Gyr ago)
and 0.12 at the present epoch. 
This relation is consistent with the recent observational results 
by Carraro, Ng and Portinari (1998).

2. Halo

The IMF of the halo is assumed to have the same form as that of the disk,
and the SFR is assumed to be constant from $t_{max} = 16$ Gyr  
to $t_{min} = 10$ Gyr (Ng {\em et al.} 1997).
	
We use the halo AMR following Ng {\em et al.} (1997):

\begin{equation}
\frac{Z-Z_{min}}{t-t_{min}} = \frac{Z(t_{max})-Z(t_{min})}{t_{max}-t_{min}}
\end{equation}

\noindent
where $Z(t_{max})=0.0004 $ and $Z(t_{min})=0.003$ are the metallicities 
at $t_{max}$ and at $t_{min}$ , respectively.
Stars in the halo will have [Fe/H] ranging from --1.7 to --0.8.
The average observed [Fe/H] in the halo is about --1.4, with
some halo stars having [Fe/H] considerably smaller than --1.7 (Larid {\em et al.} 1988).
On the other hand, the Padova evolutionary tracks are not available for 
stars with [Fe/H] $<$ --1.7.
The difference in colors between stars with [Fe/H]=--2.0 and
[Fe/H]=--3.0, at a given temperature and gravity, is usually
smaller than 0.05 mag.
So using this AMR will not introduce large shifts in the color distribution
we simulate. 

\subsection{Galactic Model}

The model consists of a halo and a disk with an  exponential scale height
growing  in time.
The shape of the halo is a de-projected de Vaucouleurs  profile 
with a flattening factor $q$ (Bahcall 1986, Table 1):

\begin{equation}
n_{halo} \propto (R/R_{0})^{7/8} \times \exp[-10.093(R/R_{0})^{1/4} + 10.093]
\times [1-0.08669/(R/R_{0})^{1/4}]
\end{equation}

\noindent
where $R_{0} = 8.5\ \rm kpc$ is the distance from the Sun to the Galactic center,
$R=(x^2+z^2/q^2)^{1/2}$, $x$ is the projected distance from the star
to the Galactic center in the Galactic plane, $z$ is the vertical
distance to the plane,
and $q$ is the axis ratio of the halo, set to be 0.8 (Reid \& Majewski 1993).

The disk is exponential in both radial and vertical directions: 

\begin{equation}
n_{disk} \propto \exp[-z/z_{0}(t)] \exp[-(x-R_{0})/h] 
\end{equation}

\noindent
with a radial scale length $h = 3.5\ \rm kpc$,
and scale height $z_{0}(t)$ that grows with time $t$

\begin{equation}
z_{0}(t) = z_{0} (1 + t / \tau_{0} )^{2/3}
\end{equation}

\noindent
where $\tau_{0} = 0.5$ Gyr and $z_{0} = 95$ pc (Rana \& Basu 1992).
The average scale height for all stars younger than the Sun is
$\sim 310$ pc, similar to the scale height of the old disk in the BS model.
This evolution of disk scale height is consistent with the results
from solving the exact dynamical evolution equations of the disk (Rana 1993).
The oldest disk stars will have a scale height of $\sim 700$ pc.
In addition, we assume that the sun is 15 pc above the Galactic plane
($z_{\odot}=15\ \rm pc$, Cohen 1995, Ng {\em et al.} 1997).
The normalization of the halo and disk population is
derived from matching the existing star counts towards the Galactic poles
(see below). 
The disk is assumed to be perfectly axisymmetric without spiral  structure.

For the Galactic extinction, we assume the dust is distributed
in a uniform layer 200 pc thick along the Galactic plane. 
We use the Burstein-Heiles (1982)  extinction map to get $A_{V}^{0}$,  the total
extinction in $V$. The extinctions in SDSS bands $A_{i}/A_{V}$ are listed
in Table 2.
The extinction in band $i$ for a star at distance $d$ and galactic latitude $b$
is:

\begin{equation}
A_{i} = \left\{ \begin{array}{ll}
	A_{V}^{0} \frac{A_{i}}{A_{V}} \frac{d\sin b + z_{\odot}}{100\ \rm pc}, &
	d\sin b + z_{\odot} < 100\ \rm pc; \\
	 A_{V}^{0} \frac{A_{i}}{A_{V}}, &
	 d\sin b + z_{\odot} \geq 100\ \rm pc 
	\end{array}
	\right.
\end{equation}

\subsection{Results of Stellar Simulations}

A simulated stellar catalog is generated as follows: 
for a certain direction on the sky, (1)  the total number 
of stars of each population is calculated according to the Galactic
model. (2) For each of the stars, its distance, age, metallicity and mass 
are generated according to the population synthesis model described
in \S 2.2. (3) The SDSS colors as well as UBVRI colors of each star
are then calculated from the stellar libraries described in \S 2.1.
(4) Galactic extinction and random 
photometric errors are added according to the prescription  in Appendix A.

The normalizations of the halo and disk populations  are determined
by fitting the total predicted star counts to  the observations
of the ESO Imaging Survey (EIS, Prandoni {\em et al.} 1998).
The EIS includes CCD observations of 1.3 deg$^2$ area towards the
South Galactic Pole in B,V and I bands.
The comparison of star counts is shown in Figure 3.
The B-V distribution at $\rm 18<V<20$ (Figure 3d) shows the bimodal distribution
of halo and disk (see also Figure 4). 
The simulation predicts $\sim 20 \%$ fewer disk stars than the observation
in this magnitude range.
A much larger area is needed to fine-tune the model parameters.
 
We show the results of the simulation in the SDSS system
towards two directions in the sky.
Figure 4 shows the color-color diagrams and CMD for a 1 deg$^2$ region
towards the North Galactic Pole (NGP). 
The halo and disk stars are represented by different symbols.
From the Figure, we notice:

1. In the $u'-g'$ vs. $g'-r'$ diagram, halo and disk stars are separated at
the blue end of the stellar locus. 
Halo stars are more metal poor, and thus bluer.
There is a well-defined blue edge of the stellar locus.
The location of this edge is  related to the age and metallicity of 
halo stars as well as the interstellar extinction.
The halo Blue Horizontal Branch stars (BHBs)
 are well-separated from the main sequence locus.
Their location on the diagram is consistent with those observed
by Krisciunas {\em et al.} (1998).
They have very blue $g'-r'$ colors.
Those stars have very similar colors to quasars at $2.5 < z < 3.0$ 
(see \S 6).
The big scatter towards the red end in this diagram is due to 
the increasing photometric errors in $u'$ band for faint red stars. 

2. In the $g'-r'$ vs. $r'-i'$ diagram, the BHBs again stand out, as they are bluer
than the main stellar locus.
Note also that $g'-r'$ remains roughly  constant
for stars with $r'-i' >0.5$, or $T_{eff} < 3500\ \rm K$.
For cool stars, absorption from molecular bands, especially from TiO,
dominate the $g'$ and $r'$ bands. Thus $g'-r'$ color is not sensitive to
the change of temperature of the star.
However, stars with different chemical composition (like carbon stars),  or with
dusty envelopes that cause internal reddening, will show up with
redder $g'-r'$ colors than normal cool stars.
We also plot the high Galactic latitude carbon stars observed by Krisciunas {\em et al.} (1998)
on this figure.
They are easily distinguishable  from normal stars.
This diagram will be useful to pick out those peculiar stars.

3. The colors, especially the IR colors, are difficult to model in cool star
atmospheres. We use different models for hot and cool stars (\S 2.1),
which  causes the slight artificial discontinuity in the $r'-i'$ vs. $i'-z'$ diagram
at $r'-i' \sim 0.7$ and $i'-z' \sim 0.7$.  

4. The CMD (Figure 4d) shows that at the bright end, most stars are disk stars,
while at the faint end, the stellar colors show the
familiar bimodal distribution: the blue peak comes from the
halo stars near the main sequence turnoff, while the red peak comes
from the disk stars.
Analysis of those diagrams from real SDSS observations will
provide important information on Galactic structure.

Figure 5 presents the color-color diagrams and CMD towards the direction ($l=0^{\circ}, b=35^{\circ}$).
This direction points toward the bulge and the Galactic plane.
It has almost the highest stellar density (for both disk and halo)
and interstellar extinction in the whole region to be surveyed by the SDSS.
Both the blue edge of the stellar locus and the position of the BHB are shifted 
to the red  by $\sim$ 0.15 mag in $u'-g'$, due to the large extinction.

Krisciunas {\em et al.} (1998) obtained photometry
of more than 1000 stars at high Galactic latitude using SDSS filters 
with the 40 inch USNO telescope.
In Figure 6 we compare their observations 
with the stellar locus from our simulations.
The solid line encloses the stellar locus from the NGP simulation. 
The stellar colors predicted by the simulation agree well with the
observations. The Krisciunas {\em et al.} (1998) data
do not go as deep as the SDSS, thus do not  include very red and faint stars.

The simulation results can be used to compare directly with the SDSS observations
and to constrain Galactic models (as we did in  Figure 3).
The star count data will be especially effective in constraining 
Galactic halo and old/thick disk models.
The distribution of the blue edge in $u'-g'$ colors depends
on the age and metallicity of the halo;
the color distribution at fainter magnitudes is
sensitive to the relative contribution from the halo and the disk (Figure 3d);
and the star counts in different directions (Figure 4,5) will be used to
constrain the shape of the halo and the scale height of the disk.
If gradients in halo/disk metallicity or age exists, 
this will also be apparent in comparing the simulation
in different directions with observations.

We do not simulate the colors of various types of 
stars in their very early and late stage of evolution,
or stars with peculiar spectral properties (like T Tau stars, late stage
of AGB stars, central stars of planetary nebulae, Ae/Be stars).
Nor do we simulate very low mass stars ($M < 0.15 M_{\odot}$) or
brown dwarfs, which may present in SDSS data especially with the inclusion
of the reddest $z'$ band. 
Many of those objects will have peculiar SDSS colors, and are interesting
in their own rights. 
However, most of them are very rare, especially at high Galactic latitude.
They are not likely to affect the studies of Galactic structure or to be 
important contaminants of quasar selection.
For the number counts  and Galactic distribution of some peculiar stars 
in various bands, see Cohen (1993).

\section{Simulation of White Dwarfs}

Hot white dwarfs have blue colors, and are one of the major
contaminants in color surveys for low redshift quasars, 
especially at the bright end (e.g. Schmidt and Green 1986). 
But by using more than two colors, one can separate the majority
of very hot white dwarfs from low redshift quasars (Koo et al. 1986).

The spatial distribution, luminosity function, and mass distribution
of white dwarfs, especially the bright, hot white dwarfs,  
have been widely studied (Weidemann 1995,
and references therein).
We simulate the colors of white dwarfs in 
the SDSS system, based on those statistics, 
in order to determine the optimal way to minimize the
white dwarf contamination in quasar selection.

1. White Dwarf Models.

The majority of white dwarfs, DA white dwarfs, 
have strong hydrogen lines, their colors 
can be modeled by a pure hydrogen atmosphere. Most of non-DA white
dwarfs are DB or DO  white dwarfs  with
He line spectra, and their colors can be modeled by a pure helium atmosphere.
There are very few white
dwarfs belonging to other spectral types. In this paper, we 
regard all non-DA white dwarfs as DB/DO white dwarfs with helium atmospheres.

Bergeron {\em et al.} (1995) have calculated H and He white dwarf
atmosphere models. Combining these with a detailed
evolutionary cooling sequence (Wood 1995), they  obtain for each model the absolute
magnitude, mass and age. Dr. Bergeron kindly calculated the SDSS
colors of those models, which we use here as the basis of our
simulation.

2. Spatial Distribution.

We only consider disk white dwarfs, as the density of halo white
dwarfs is about two orders of magnitude smaller 
(Liebert {\em et al.} 1988) and
they are much farther away. 
The disk white dwarfs have a spatial distribution
similar to that of old disk stars, with a scale height of about 275 pc
(Boyle 1989).
Most of the white dwarfs bright enough to be observed by the SDSS are very nearby,
so the disk can be regarded as 
plane parallel; the density of white dwarfs thus does not depend on
the Galactic longitude.

3. Luminosity Function and Mass Distribution.

Liebert {\em et al.} (1988) have determined the local luminosity function 
of DA and non-DA white dwarfs (see also Oswalt {\em et al.} 1996).
The LF is determined down to $M_{bol}=15.75$.
White dwarfs with $M_{bol} > 14$ have $T_{eff}<6000K$, with colors
similar to normal stars. They are not important for quasar selection.
Bergeron {\em et al.} (1992) have determined
the mass distribution of hot white dwarfs, 
$M= 0.56 \pm 0.14 M_{\odot}$. We use those results in our simulation,
assuming  that the mass distribution is Gaussian and 
is independent of the luminosity. 

To generate the simulated white dwarf catalog for a specific direction
on the sky, we (1) calculate the total number of DA and non-DA white
dwarfs based on their local LF and spatial distribution; (2) for
each white dwarf, randomly choose  its luminosity and mass from the 
LF and mass distribution; (3) knowing the mass and luminosity, 
determine the SDSS magnitudes from Bergeron's atmosphere models;
(4) finally, add the Galactic extinction  
and photometric errors as was done for stars (\S 2.2).

Figure 7 shows the number--magnitude relation of white dwarfs 
towards $b=60^{\circ}$ in $g'$.
In the figure, we also plot the number-magnitude
relation for quasars. 
For $g'<17$, the density of white dwarfs is higher than quasars
at $b=60^{\circ}$. White dwarfs are major contaminants of the 
bright quasars.
In Figure 8, we plot the white dwarf density as a function of Galactic
latitude (not including the effect of Galactic extinction). 
The number density of white dwarfs is much higher towards the plane.

Figure 9 plots the color-color diagrams of a simulation of white
dwarfs (solid circles) for an area of 20 deg$^2$ centered at $b=60^{\circ}$.
Also shown are the synthetic SDSS colors (open circles) 
from the spectrophotometric atlas
of white dwarfs of Greenstein \& Liebert (1989), 
calculated by Lenz {\em et al.} (1998).
The atlas consists of white dwarfs of all spectral types and temperatures.
Their relative density does not represent reality.
It is evident that the model colors of white dwarfs agree with the synthetic
colors of real observations quite well. 
For hotter white dwarfs, DAs and non-DAs follow slightly different cooling tracks
on the color-color diagram, especially in the $u'-g'$ vs. $g'-r'$ plot, since
DA white dwarfs have a Balmer jump (although weaker than normal stars due
to the very high gravity). When the white dwarf gets cooler, the colors
of DA and non-DA white dwarfs become more similar to each other, and  to normal stars.

We will examine the distribution of white dwarfs in color space and
how to separate them from quasars in more detail in \S 6.

\section{Simulation of Quasars}

Simulated catalogs of quasars in the SDSS system will be used to 
determine the color selection criteria of quasars (\S 6).
The quasar selection function, critical to the determination
of the quasar LF, also requires detailed modeling of the distribution of
quasars in color space (e.g. Warren {\em et al.} 1994).
We generate the simulated quasar catalog based on the colors
calculated from a quasar spectral model (\S 4.1) 
and the observed evolution of the quasar LF (\S 4.2). 

\subsection{Quasar Spectral Model}

A quasar spectrum consists of three components: (1) continuum,
which for a limited range in optical and UV, can be approximated
by a power law, $f_{\nu} \propto \nu^{-\alpha}$;
(2) emission lines; 
(3) line and continuum absorptions.

In the simulation, we assume that the power law index $\alpha$
has a Gaussian distribution $\alpha = 0.5 \pm 0.3$ (Francis 1996).
The distribution of the indices may in fact not be Gaussian, and may also
be a function of redshift (or wavelength). For example, 
Sargent, Steidel and Boksenberg (1989) determined from
a sample of high-redshift quasars $\alpha = 0.78 \pm 0.20$.
A difference of 0.4 in the power law index  will result in a difference in
broad-band colors of $\sim 0.1$ mag (Fan \& Chen 1994). 
Small changes in the power law indices will therefore not significantly change
quasar colors.

The emission line flux ratios of 16 major lines in quasar spectra are taken from the
statistics of the sample of Wilkes (1986).
This sample consists of radio-selected quasars, and therefore
does not have a priori bias in the optical emission line statistics.
We also include 20 weaker lines and FeII features (those not measured in
Wilkes 1986), using the line ratios
from Francis {\em et al.} (1991) based on the Large Bright Quasar Survey (LBQS) sample (Hewett {\em et al.} 1991 and references therein).
The equivalent width (EW) of $\rm Ly\ \alpha$ is assumed to follow
a Gaussian distribution of 65 \AA\ $\pm $ 34 \AA\ (Wilkes 1986).
The line ratios are assumed to be the same for all quasars, with only 
the strength of $\rm Ly\ \alpha$ varying.
For broad lines in quasar spectra, we assume the FWHM is 5000 $\rm km\ s^{-1}$ 
with a Gaussian shape.
For narrow forbidden lines, FWHM = 1000 $\rm km\ s^{-1}$.
The width of FeII features are taken from Francis {\em et al.} (1991).
Different assumptions on the width of the lines have little effect on
the broad-band colors.

High-redshift quasars have complicated absorption systems. 
Those systems are the major factor for the color evolution of quasars with redshift.
Various people have simulated the absorption spectrum and its
effects on broad band colors at high redshift (e.g. Giallongo 
\& Trevese 1990, M\o ller \& Jakobsen 1990, Fan \& Chen 1993, 
Warren {\em et al.} 1994). We follow the prescriptions by
Fan \& Chen (1993) (similar to those used by Warren {\em et al.} 1994)
to simulate the absorption spectrum caused by intervening HI
absorbers ($\rm Ly\ \alpha$ forest, damped $\rm Ly\ \alpha$ system
and Lyman Limit Systems, LLSs). The metal absorption lines have very
small equivalent width and do not affect colors significantly.
Details of the absorption simulation and parameters used are summarized in
Appendix B. 

In Figure 1, we give an example of a simulated quasar spectrum
at $z=2.8$, overplotted on the filter response curves of the SDSS
system. The spectrum has a resolution of 10 \AA. Note that at this redshift,
Ly $\alpha$ emission is in the $g'$ band, and the $\rm Ly\ \alpha$ forest
lines absorb a large fraction of the light in the $u'$ band.

\subsection{Quasar Evolution}

The quasar LF has very strong redshift evolution. 
Several analytical models for the quasar LF have been proposed 
(e.g. Schmidt \& Green 1983, Boyle, Shanks \& Peterson 1988,
Warren {\em et al.} 1994, Pei 1995).
Pei (1995) fit the evolution of quasars over the range
$0 < z < 4.5$ with a pure luminosity evolution model.
He used the empirically determined quasar luminosity
function from the samples of Hartwick \& Schade (1990) ($ z < 2.2$) and 
Warren {\em et al.} (1994) ($2.2 < z < 4.5$).
The shape of the luminosity function is described as a 
double power law or an exponential $L^{1/4}$ law, while the characteristic
luminosity of the quasar evolves with redshift. 
For the double power law model, 

\begin{equation}
\Phi(M,z) = \frac{\Phi^{\ast}}{10^{0.4[M-M^{\ast}(z)](\beta_{l}+1)}
			+10^{0.4[M-M^{\ast}(z)](\beta_{h}+1)}}
\end{equation}

\noindent
where  $\beta_{l}$ and $\beta_{h}$ are the indices of the power laws
and $M^{\ast}(z)$ describes the luminosity evolution:

\begin{equation}
M^{\ast}(z) = M^{\ast}_{0} + 2.5 (1 - \alpha) \log (1+z) +
1.086 \frac{(z-z^{\ast})^{2}}{2\sigma_{z}^2} 
\end{equation}

\noindent
$\alpha$ is the average power law index of the quasar continuum.
In this form of luminosity evolution, the characteristic luminosity
evolves as a power law of $\sim (1+z)^3$ at low redshift, 
and reaches its maximum at $\sim z^{\ast}$, 
then decays as a Gaussian towards higher redshift.
The absolute magnitude is evaluated in $M_{B}$ (including K-correction), 
and $\Phi$ has  the unit of 
$\rm Mpc^{-3}\ mag^{-1}$ .
We use ($h, q_{0}, \alpha) = (0.5, 0.5, 0.5)$ throughout the paper.
For these parameters, the best fit luminosity function is given
by: 
$\beta_{l}=-1.64$, $\beta_{h}=-3.52$, $M^{\ast}_{0}=-27.10$,
$z^{\ast} = 2.75$, $\sigma_{z} = 0.93$, and $\Phi^{\ast} = 8.22 \times 10^{-7}\
\rm Mpc^{-3}\ mag^{-1}$. 

The sample in Warren {\em et al.} (1994) includes only one quasar
with $z>4$.
and   the  slope of decay for the fit in Pei (1995) is not well-determined.
Indeed, using this fit  predicts much fewer quasars at $z>4$ than are
found in  the surveys of Schneider, Schmidt \& Gunn (1991a,1994, SSG thereafter), Kennefick {\em et al.} (1997) and
Irwin {\em et al.} (1991).
So in the simulation we use the fit of Schmidt, Schneider \& Gunn (1995)
for the quasar LF at $z>4.0$:

\begin{equation}
\log \Phi(z, <M_{B}) = -6.835 - 0.43(z-3) + 0.748 (M_{B} + 26)
\end{equation}

\noindent 
or,
\begin{equation}
\Phi_{hz}(M,z) = \frac{\Phi_{hz}^{\ast}}{10^{0.4[M-M_{hz}^{\ast}(z)](\beta_{hz}+1)}}
\end{equation}

\noindent
where $M_{hz}^{\ast}(z)= -27.72 + 0.57z$, $\beta_{hz}=-2.87$ and 
$\Phi_{hz}^{\ast}=2.42\times 10^{-7} \rm Mpc^{-3}\ mag^{-1}$. 

The simulated quasar catalog is generated as following:
(1) for a specific area on the sky, the  total number of
quasars for the low ($z<4$)  and high redshift ($z>4$) subsamples are calculated
from eqs. (9) and (11), respectively,
and the redshift and absolute magnitude (in $M_{B}$) 
of each quasar are derived from their distribution functions.
(2) For each quasar, the continuum and emission + absorption  spectra 
are determined.
The SDSS colors are calculated from the spectrum.
(3) Finally, the Galactic extinction and photometric errors 
are added.

The number-magnitude relation of quasars in $i'$ band is shown in Figure 10. 
We compare the predicted numbers of quasars at low and high redshift
with observed numbers (normalized to the survey area of 10000 $\rm deg^{2}$).
For $z < 4.0$, our numbers agree well with the results of Crampton 
{\em et al.} (1987, dotted line). To transfer $B$ to SDSS $i'$, 
we assume $i' = B - 0.5$, appropriate for a quasars with $z < 2.5$.
For $z > 4.0$, we compare our predictions with the observations
of (1) SSG, who found 10 quasars with $z>4$ in 61  $\rm deg^{2}$.
We  assume the limiting magnitude of SSG is $r \sim 21$, and
$i' \sim r - 0.3$ for quasars with $z \sim 4.3$ ( Fig. 12c); 
(2) the DPOSSII survey (Kennefick {\em et al.} 1996), 
which found 9 quasars with $z>4$ in
681 $\rm deg^{2}$ with limiting magnitude $r < 19.6$, assuming
an average efficiency of 50 \% as estimated in the paper;
(3) the APM survey (Irwin {\em et al.} 1991, Storrie-Lombardi {\em et al.} 1996), 
which found 27 quasars with $z>4$ and
$R < 19$ in 2500 $\rm deg^{2}$.
Our predicted numbers for $z>4$ agree well with those observations.
In the whole SDSS survey one expects to find $\sim 100,000$ quasars to the
spectroscopic survey limit of $i' \sim 19$. 
Figure 11 shows the predicted numbers of quasars  
as a function of redshift for different magnitude limits.

\subsection{Evolution of Quasar Colors with Redshift}

Figure 12 shows the evolution of the median and scatter of quasar colors
with redshift in the SDSS bands, calculated from
100 Monte-Carlo realizations in each bin of 0.1 in redshift. 
From this plot, we notice the following: 

(1) For $z<2.0$, the $u'-g'$ color ( Fig.12a ) of quasars remains smaller than
0.3.
In this redshift range, quasar colors are dominated by the power-law continuum.
The colors of a pure power-law spectrum are independent of
redshift for a given index $\alpha$.
All quasar emission lines other than  Ly $\alpha$
have equivalent widths smaller than 100 \AA. 
and thus only have small effects on broad-band colors. 
The small change in the mean colors at
low redshift is caused by those emission lines passing through the $u'$ and $g'$
bands. 
At $z \sim 2.0$, $\rm Ly\ \alpha$ emission enters the $u'$ band,
making the $u'-g'$ color bluer. When  $\rm Ly\ \alpha$ emission
moves from $u'$ to $g'$ at $z \sim 2.5$, the $u'-g'$ color gets redder.

(2) For $z>2.5$, the absorption systems, first the $\rm Ly\ \alpha$ forest,
then the LLSs, enter the $u'$ band. They absorb most of the continuum
radiation in the $u'$ band, and the $u'-g'$ color reddens quickly with redshift.
Meanwhile, the random distribution of the absorption systems, 
especially, the presence or absence of LLS, optically thick to continuum,  in the $u'$ band,
causes very large scatter of color for a given redshift.
The number density of LLS grows with redshift (see Appendix B); for
$z>3.7$,  very few quasars still have detectable $u'$ flux. 

(3) The evolution in  redder colors ($g'-r'$, $r'-i'$, $i'-z'$, Fig.12b-d 
)
has similar behavior to that of $u'-g'$,
only shifted to  higher redshift. LLSs dominate the $g'$, $r'$ and $i'$ bands from
$z= 4.0, 5.0$ and  6.0, respectively.
The major difference of the quasar color evolution of those 
colors from that of $u'-g'$ is that there is no obvious dip at the redshift
when Ly $\alpha$ emission is in the bluer band, as it is when the
emission is in $u'$ at $z \sim 2.0$.
This is because for quasars with $z >4$, Ly $\alpha$ forest lines absorb more than
half of the continuum blueward of Ly $\alpha$ emission,
so the increase of flux from Ly $\alpha$ emission in  a certain band is cancelled by 
absorption immediately blueward of  it.

Figure 13 compares the simulation results with observations of 
quasars in SDSS colors. The simulation is for 
an area of 10 deg$^2$; only quasars with $i'<19.5$
are plotted. 
There are very few quasars at $z>3.5$. To generate enough
simulated quasars for comparison, we assume the quasar  LF
to be constant at $z>3.5$, 
instead of the evolving LF described above.  
The observations are taken from test observations of
known quasars in the SDSS bands by Richards {\em et al.} (1997),
Newberg {\em et al.} (1999) and Fan {\em et al.} (1999).
The simulated colors agree with the observations very well,
with few exceptions, where the observed quasars have Broad Absorption Lines
(BALs), not being simulated in this work.
In Figure 13d, we plot the median and scatter of $i'$ magnitude for a 
quasar with $M_{B} = -27$, a very bright quasar similar to 3C273,
as a function of redshift. 
SDSS could easily detect such a quasar at $z>5.0$, if there were any
in the universe. 
Discovery of such quasars would a have profound impact on the studies of
the structure formation and early galaxy evolution.

In \S 6, we will compare the quasar colors with other types of
stellar objects, and investigate how to separate quasars from
other objects in color space in detail.

\section{Simulation of Compact Emission Line Galaxies}

Some  Compact Emission Line Galaxies (CELGs) have power law continua and strong 
emission lines, and thus their colors are similar to those of low-redshift
quasars.
Some CELGs have very strong narrow emission lines, with 
equivalent width exceeding 1000 \AA. 
Those strong emission lines affect the colors greatly, and this effect
can  change with redshift as the emission lines pass through different
bands. 
The emission spectra of CELGs differ greatly in ionization and excitation levels.
Their  continua can be  powered by starburst activity or a 
central black hole. 
Their luminosity function is not well-established (Salzer 1989), 
and little is known about their redshift evolution (Koo {\em et al.} 1994).
Furthermore, the possibility that those objects will be identified
as stellar objects in the SDSS is a function of both the observing
condition and the surface brightness profiles of those objects.
It is therefore very difficult to simulate the distribution 
of CELGs in the SDSS color space.
Instead of attempting to develop a population synthesis  model for CELGs
(as we did for stars and quasars), 
we construct the  simulated catalog by directly using the spectroscopic
data of the Schneider-Schmidt-Gunn (SSG, 1994) survey.  
We first calculate the SDSS magnitudes of the SSG galaxies according
to their line fluxes (Horowitz 1994) and the continuum slope distribution
of McQuade {\em et al.} (1995). 
In constructing the catalog, we randomly select galaxies from the SSG 
sample (for $r' < 19$, where they are complete, and some extrapolation
beyond that), and assign the angular size of each object according
to the size distribution of Salzer {\em et al.} (1989).

\subsection{CELG Model}

1. SSG Sample.

We first want to define a flux-limited, uniformly-selected subsample from the SSG
sample.
The SSG survey selects emission line object candidates based
on Palomar 4-shooter grism spectra.
Their emission line galaxy sample goes deep (close to the magnitude
limit of the SDSS spectroscopic survey for quasars, $i' \sim 19$)
and has large sky coverage (61.4 sq. deg).
Horowitz (1994) analyzed the spectral properties of the SSG sample.
He measured the line flux and EW of $\rm H\beta + [OIII]$,
as well as line ratios of $\rm H\alpha$, [NII], [SII] and  [OII]
for all the SSG galaxies that were originally selected by their
 $\rm H\beta + [OIII]$ emission on the grism spectra.
Horowitz (1994) also classified the spectral type of the CELG
to be one of five types: Seyfert 1, Seyfert 2, Starburst,
HII galaxy and LINER.
This information is the basis of our simulation.

Horowitz (1994) doesn't include galaxies detected only by $\rm H\alpha$ from the grism. 
They make up about 30\% of all the SSG galaxies.
But they are mostly of very low excitation (otherwise they would have
been detected by their $\rm H\beta + [OIII]$ emission), 
an indication of low activity, thus
unlikely to have non-stellar colors, and therefore not a serious
problem for the quasar selection.
Furthermore, they are at low redshift (in order for $\rm H\alpha$
to be within the range of grism spectra), and thus have bigger sizes, making
them unlikely to be classified as stellar objects by the SDSS.
We ignore those galaxies in the current simulations.

SSG (1994) select objects according to  two criteria:  
(1) equivalent width of $\rm H\beta + [OIII]$ greater than 50 \AA;
and (2) line flux greater than about
 $\rm 10^{-14.1} erg \hspace{2mm} s^{-1} cm^{-2} \AA^{-1}$.
Transferring these constraints to continuum flux,
this sample is ``complete'' for all galaxies with equivalent width
greater than 50 \AA\ and $m^{con}_{AB} < 19$.
For fainter galaxies, this sample is only complete for galaxies with even stronger lines.
Therefore, we select a  ``complete'' subsample of 167 galaxies with $r' < 19$.
and refer it as SSG19. 

The redshift distribution of the SSG sample is fitted by a $\Gamma$ distribution,
$N(z)  \propto z \exp(-z/z_{0})$ where $z_{0}$ = 0.08,
and will be used in the simulation to generate fainter CELGs.
The number count vs. magnitude relation is consistent with $N(m) \propto 10^{0.4m}$.
For $r'=19$, it has a surface density of 2.7 per deg$^{2}$.

2. McQuade {\em et al.} sample

McQuade {\em et al.} (1995) provide  UV to optical spectrophotometry of a sample
of 31 emission line galaxies.
From this sample, we calculate the continuum slope of different
types of CELGs. We assume the continuum of CELGs to be power-law
in the relevant wavelength range, the power-law indices of
each type are :
Seyfert 1: $\alpha = -1 \pm 0.5$; Seyfert 2: $\alpha = -2 \pm 0.5$;
Starburst: $\alpha = -1.5 \pm 0.5$; HII galaxy : $\alpha = -1 \pm 0.5$; LINER :  $\alpha = -2 \pm 0.5$.

The spectrum of an emission line galaxy is assumed to be a power law
continuum plus 16 major UV and optical emission lines.
For each galaxy in the SSG19 sample, we randomly select
a continuum slope based on the statistics above, assuming Gaussian
distribution.
If a line ratio is not measured in Horowitz (1994),
we will assume a flux ratio (relative to $\rm H \beta$)
based on the measurements of McQuade {\em et al.}, according to its type:
for Seyfert, we use the flux ratios of NGC1068, 
for Starburst/LINER, NGC 6052, and for HII galaxy, UGC9560.
The SDSS magnitudes of all the SSG19 galaxies are calculated.

3. Size Distribution

Salzer {\em et al.} (1989) studied the properties of a deep, 
complete sample of emission line galaxies from the University of Michigan survey.
The isophotal magnitude and isophotal radius are determined for
each galaxy.
Using their sample, we get a relation between the absolute  magnitude
and isophotal radius:

\begin{equation}
  \log D_{25}=(-2.424\pm0.093)-(0.183\pm0.005)M_{B}
\end{equation}

The r.m.s. scatter of $ \log D_{25}$ around this relation is 0.119.
This relation is plotted in Figure 14. 
Different spectral types are represented by different symbols.
It shows no strong selection effect against
low surface brightness galaxies towards fainter magnitude.
Assuming an exponential disk, one can derive the scale length 
and characteristic surface brightness $R_{e}$ and $\mu_{0}$.
The distribution of $\mu_{0}$ is roughly a Gaussian, peaking at
 $\rm \sim 20.6\ mag\ arcsec^{-2}$, 
1 mag brighter than the Freeman (1979) central surface brightness of disks.
In the simulation, the $ D_{25}$ of a CELG is calculated from eq (12),
with  a Gaussian scatter of 0.119 in logarithm. 
$R_{e}$ is then calculated assuming a pure exponential disk. 
The SDSS imaging observations will be taken only under seeing $\lesssim 1''$. 
Under such observing conditions, the photometric pipeline will be able to 
resolve galaxies with $R_{e} > 0.5''$ at $r'<19$.
Therefore, the simulated catalog only includes those galaxies
with $R_{e} < 0.5''$.
About 50 \% of the CELGs at $r'< 19$ in the simulated catalog 
have $R_{e} < 0.5''$.

\subsection{Colors of CELGs}

The simulated catalog is constructed separately for two subsamples : 
the brighter sample where SSG is roughly complete (SSG19, $r' < 19$ ); 
and the fainter sample.

 1. Brighter sample : after specifying the size of the area, 
(1) the average number of galaxies with $r'<19$ in this area  is calculated. 
(2) The actual number $N$ of galaxies
is selected randomly from a Possion distribution. 
(3) $N$ galaxies from the SSG19 sample are randomly chosen 
and the SDSS magnitudes are calculated according
to Horowitz's measurement and the spectral model described above.
Note that this brighter sample is based on the SSG sample;
in particular, there is no  assumption about the evolution or luminosity function
of these objects.
The brighter sample is as deep as the SDSS spectroscopic survey for quasars.

 2. Fainter sample : after specifying the size of the area, 
(1) the average number of galaxies 
with $19 < r' < 24 $ is calculated 
assuming  $N(m) \propto 10^{0.4m}$ and the normalization
of SSG19. 
(2) The actual number $N$ in the fainter sample is selected randomly from a Possion distribution.
(3) For each of the galaxies, the $r'$ magnitude and  redshift are
drawn randomly from the distribution of SSG galaxies 
in \S 5.1. 
(4) We randomly pick $N$ galaxies from the SSG sample. For each of the galaxies,
we use  its type, the EW of $\rm H\beta + O[III]$, and line ratios, 
and  determine the  continuum slope and line ratios of weaker lines from
the spectral model described above according to its spectral type.
We calculate the SDSS magnitudes from
the spectrum.

 We calculate the angular size of the galaxy as described above,
rejecting objects with $R_{e} < 0.5''$. 
Finally, we add the Galactic extinction and photometric errors to each of the galaxies.

In Figure 15, we present the color-color diagrams and
size distribution of a simulation of CELGs in 10 deg$^2$.
We also compare the simulation with
the colors of CELGs found by Hall {\em et al. } (1996,
catalogs given in Osmer {\em et al.} 1998), transferred to the
SDSS colors using the relations in F96.
We only plot the CELGs with $z<0.25$ in Hall {\em et al.}'s sample
(their survey is much deeper than the SDSS quasar survey).
From the Figure, we notice:

1. The SDSS colors of most of the emission line galaxies are dominated by
the continuum. This is simply due to the large bandwidth of
SDSS filters. An emission line EW of 100 \AA\ will
only have about 0.1 mag effect on the magnitude at most (if the line
is at the center of the filter).
Half of the galaxies in SSG19 have $\rm EW(\rm H\beta + [OIII]) < 100 \AA$.
The obvious locus of emission line
galaxies in Fig.15 is the color of this power law continuum.

2. The colors of CELGs are on average redder than those of quasars,
due to their redder continua.
The accurate modeling of the locations of those galaxies in color space
critically depends the choice of power law distribution.
Hall {\em et al.} (1996) conclude that putting a limit on $B-V < 0.6$ will
eliminate half of the CELGs from the quasar candidates.
Our simulation also includes some galaxies redder than those of Hall {\em et al.}'s.
They have colors similar to normal late type galaxies and will not
be picked up by Hall {\em et al.}'s color selection criteria.

3. Some CELGs have very strong emission lines and almost no continuum,
leading to enormous EW (some of the SSG galaxies 
have emission lines with EW exceeding 6000 \AA),
which results in a color change of several magnitudes from the power law. 
The redshift effect will make their colors even more peculiar.
Those CELGs are located in parts of color space very far from normal stars and
quasars. 

The number-magnitude relation of CELGs in $i'$ band is presented in Figure 16.
We compare it with the number count of quasars (Figure 10). 
The number density of CELGs is 20 -- 25\% of the number density of quasars
for a large magnitude range (see also Koo {\em et al.} 1994). 
However, since colors of CELGs  are redder than those of quasars,
rejecting sources with $g'- r'>0.35$ will eliminate about 60\% of CELGs while 
retain most of the quasars. We will further discuss the separation of
quasars and CELGs in \S 6.

\section{Discussion}

\subsection{Distribution of Stellar Object in the Color Space}

In this subsection, we summarize the results of the simulations
described in the previous sections,
and show how different types of stellar objects occupy
different regions of the color space.
A collection of the simulated catalogs, for quasars, CELGs,
and for stars and white dwarfs towards several directions
on the sky, is available at : \texttt{http://www.astro.princeton.edu/$^{\sim}$fan/sdss\_simu.html}.

Figure 17 shows a 10 deg$^2$ simulation of all four kinds
of stellar objects toward the NGP, with $i' < 19.5$,
in $u'-g'$, $g'-r'$, $r'-i'$ color space.
In Figure 18, we show the approximate locus of each 
type of object on color-color diagrams, estimated from Figure 17.
From Figs 17 and 18, we can study the relative distribution of stellar objects
in color space:

1. Bright and hot white dwarfs occupy the bluest corner of
color space. First, they are hotter and bluer than
normal stars. Second, they have similar $u'-g'$  colors to quasars,
but with bluer $g'-r'$,$r'-i'$ and $i'-z'$ colors. 
For hot white dwarfs ($T_{eff} > 10000\ \rm K$) , their spectra
are already in the Rayleigh-Jeans part of the black-body
curve at $\lambda > 5000 \rm \AA$, i.e., $f_{\nu} \propto \nu^{2}$,
much bluer than those of quasars or AGNs with $f_{\nu} \propto \nu^{-0.5}$.
This property can be used to separate hot white dwarfs from quasars.

2. Low redshift quasars ($z < 2.0$) have blue $u'-g'$ colors.
They are well-separated from stars and white dwarfs in color space,
and may only be confused with some CELGs.
The position of quasars at $z<2.0$ changes little with redshift,
due to the dominant power-law continuum. 
For $z > 2$, the $u'-g'$ color becomes increasingly red.

3. Quasars with $2.5 < z < 3.0$, white dwarfs with $T_{eff} \sim$ 7000 K,
blue halo stars and some CELGs intersect in the region of
color space with $u'-g' \sim 0.8$, $g'-r' \sim 0.3$ and
$r'-i'$, $i'-z' \sim 0$. 
For quasar at this redshift range, (a) the Ly $\alpha$ emission is 
redshifted to $\sim 4500$ \AA, (b) the continuum radiation blueward
of Ly $\alpha$ emission is absorbed by Ly $\alpha$ forest lines and
LLSs, causing a break at $\sim 4000$ \AA\ in the SED, and (c) 
the continuum redward of  Ly $\alpha$ emission gradually declines
towards longer wavelength. 
These three features have very similar wavelengths to the features
in a stellar spectrum of  $T_{eff} \sim$ 7000 K: (a) the flux peak
at $\sim 4500$ \AA, (b) the Balmer jump and Balmer absorption lines 
at $3500 - 4000$ \AA, and (c) the gradual decline of flux at longer
wavelengths. 
These coincidences cause the broad-band colors of these objects to be very
similar (see also Fig 1). 
This is the most difficult redshift range for quasar selection.
Furthermore, since both the stellar populations and Galactic
extinction change with Galactic position, the detailed distribution
of stars and white dwarfs in color space also change,
making the quasar selection even more complicated.

4. For $z>3$, quasars have very red colors because of the
intervening absorption systems in their spectra.
They are well separated from other kinds of stellar objects in
color space.

5. CELGs have a big scatter in their distribution in color space.
Most of them have redder colors than quasars due to their redder continua.
Those with AGNs have similar colors to  quasars. Some HII galaxies,
with very strong optical emission lines dominate their spectra,	 
can have very peculiar colors. 

\subsection{The ``Fundamental Plane'' in Color Space}

Newberg {\em et al.} (1999) point out that in their observations,
stars, galaxies and low redshift quasars are distributed approximately
in the same ``fundamental plane'' in SDSS color space.
The stellar locus, in particular, forms a ribbon-like structure in
color space.
Our simulations further demonstrate this point. 
Using the algorithm developed in Newberg \& Yanny (1997), we
fit a set of stellar locus points to our stellar distribution
in the simulated color space of ($u'-g', g'-r', r'-i'$). 
This locus is shown in Figure 18.
The distribution of stars perpendicular to this locus 
is then fitted by an ellipse.
We choose a set of axes :

\begin{equation}
\left\{ \begin{array}{lll}
		c1 & = & 0.95(u'-g') + 0.31(g'-r') + 0.11(r'-i') \\
		c2 & = & 0.07(u'-g') - 0.49(g'-r') + 0.87(r'-i') \\
		c3 & = & -0.39(u'-g') + 0.79(g'-r') + 0.47(r'-i') 
	\end{array}
\right.
\end{equation}

\noindent
$c1$ is along the average direction of the stellar locus for $T_{eff} > 4000$ K (in Fig.18);
$c2$ is along the major axis of the fitted ellipse perpendicular to
the stellar locus; and $c3$ is along the minor axis of the ellipse.

In Figure 19, we plot the NGP simulation in the projections of
$c1$ vs. $c2$ and $c1$ vs. $c3$, i.e., the edge-on and
face-on view of the ``fundamental plane'' on which most of
the stars are distributed.
We show the reddening vectors on these projections.
We also plot the tracks of black body spectra  (dotted line, from
2000 K to 30000 K) and power law spectra (solid line, $\alpha = 0.0 - 2.0$)
on those projections.

From Figure 19(a), it is evident that: 
(1) normal stars, white dwarfs, as
well as quasars and most of the CELGs are distributed in a plane,
with $c2 = 0.07(u'-g') - 0.49(g'-r') + 0.87(r'-i') \sim 0.0 \pm 0.1$.
Note that $c2$ has almost no contribution from $u'$.
For a black body spectrum, $c2 \sim 0$ for all temperatures.
The fact that  all but the coolest stars are 
distributed on this plane merely reflects the fact that their SEDs
are close to black-body except in the $u'$ band, which covers 
the Balmer jump for hotter stars, and the metal line blending in UV
for cooler stars.
(2) For M stars ($T_{eff} < $ 3500 K), the optical spectra are dominated
by molecular absorption, and thus have big deviations from a black body spectrum.
They therefore depart from the ``fundamental plane''.
(3) The colors of quasars and CELGs are dominated by the power law continuum.
They have $c2 \sim 0.05 - 0.10$.
(4) The reddening vector points almost entirely in the $c1$ direction.
Galactic extinction can only move objects within the plane.
The blue edge of the stellar distribution in the $c1\ \rm vs.\ c2$
projection depends on the stellar population as well as reddening.
The points above explains the existence of such a ``fundamental plane''
in color space.

Figure 19(b) shows the face-on view of the ``fundamental plane''.
On this plot, different types of objects are well separated.
For a fixed temperature, changes in stellar metallicity, gravity and
reddening will result in different positions on the $c1 - c3$ plane.
The BHBs stand out clearly from the stellar locus.
They are also located in different positions from the white dwarfs with
similar temperature ($c1 \sim 1.0 $ for BHBs, and $c1 \sim 0.3$ for
white dwarfs with similar temperature), due to the large difference in
their surface gravities and therefore different strength of the Balmer
jump. 

\subsection{The Quasar Selection Problem}

It is not the goal of the current paper to discuss the quasar selection
algorithm of the SDSS in great detail. 
Aspects of it are discussed in Newberg \& Yanny (1997).
The final algorithm will be established during the SDSS test year.
However, we intend to show through simulations the separation of 
quasars and other objects in color space, following the discussion
in \S 6.1.

In Figure 21, we plot the median distance and its 68\% scatter
of quasars from the stellar locus (see \S 6.1) in color space
as a function of redshift.
We show both the distances based on three colors ($u'-g', g'-r', r'-i'$)
and based on the two redder colors.
It demonstrates two problems with the separation of quasars from stars with colors.
(1) At redshift 2.5 -- 3.0, quasars move very close
to the stellar locus in color space.
Some of the quasars have almost identical colors in all bands to  
blue halo stars as we saw above. 
Any quasar selection based on color will miss a fraction of all quasars in this redshift range or will suffer a large contamination from stars.
Note that this redshift range is also where the quasar LF peaks (\S 4.2) .
The evolution of the quasar population in this redshift range is important
to cosmology. But we will unavoidably need a big selection correction here.
Detailed modeling of the quasar population, as we described in \S 4, 
is crucial in determining the quasar selection function (see also
Warren {\em et al.} 1994).
(2) For $z > 3.3$, quasars get increasingly redder in $u'-g'$.
They are well separated from the stellar locus.
However, they will have very faint $u'$ detections, or in most cases
for $z > 4$, will be completely absent from the $u'$ images.
One has to use the lower limit in $u' - g'$ to separate quasars and stars. 
Quasars with $z < 4.2$ can have very similar $g'-r'$, and $r'-i'$ colors to stars.
Quasars with very high redshift ($z > 4.2$) can be separated
from stars based on $g'-r'$ and $r'-i'$ colors alone.

White dwarfs, BHBs and CELGs are also rare (compared to normal stars) and
are well separated from the stellar locus.
Simply by selecting objects as outliers from the stellar locus (e.g. Newberg \&
Yanny 1997) will include those contaminants in the quasar candidate list.
White dwarfs will outnumber quasars for $r' < 17$, while 
CELGs have a number density about 25\% of  that of quasars at the faint end
(Koo {\em et al.} 1995).
In Figure 20, we show the separation of quasars in the redshift
range $2.2 < z < 3.2$
and stars/white dwarfs/CELGs.
As discussed in \S 6.1, hot white dwarfs (with $T_{eff} >$ 8000 K)
have similar $u'-g'$ but bluer $g'-r'$ color than quasars;
a cut at $g' - r' > -0.20$ will eliminate most of them.
At $r' < 17$, most of  the white dwarfs are hot and have $g' - r' < -0.20$.
At $r' < 19.5$,  there are more cooler white dwarfs, and $\sim$ 60\% of white dwarfs 
can be eliminated in this way.
Most of the CELGs with HII or starburst spectra will have redder continua
than those of quasars or Seyfert galaxies. 
By rejecting  objects with $g'-r' > 0.35$, we
exclude $\sim$ 60\% of CELGs (see also Figure 16).

\section{Summary}

 We present in this paper a  simulation of the 
color-magnitude distribution of the stellar sources in the SDSS
filter system. We simulate the distributions of Galactic stars,
white dwarfs, quasars and compact emission line galaxies that
the SDSS photometric pipelines will pick as stellar sources by morphology.
We do not simulate the colors of normal galaxies
that may appear stellar for the SDSS.
We also do not simulate various kinds of peculiar stars (see \S 3.4).
With these exceptions aside, we have covered the major classes of
stellar objects that the SDSS will observe at high Galactic latitudes.
In this paper, we present the models of the spatial and spectral distributions
for each type of object. We also add realistic photometric 
errors  and Galactic extinction in the simulation. 
Simulated catalogs of SDSS observations towards several directions
on the sky are available at  \texttt{http://www.astro.princeton.edu/$^{\sim}$fan/sdss\_simu.html}.

 We simulate the distribution of stars based on (1) evolutionary stellar
population synthesis models of the disk and halo populations, 
using the Padova stellar isochrones and Lejeune {\em et al.}'s (1997a,b)
stellar atmosphere libraries; (2) a Bahcall-Soneira  (1984) type Galactic
model, including a disk whose height grows with stellar age, and
a flattened stellar halo.
The disk stars  are separated from
the halo stars in color-magnitude diagrams as well as in color-color space. 
The simulation tools will be  very useful for constraining Galactic star
count models as well as for finding peculiar stars.

 We simulate the distribution of white dwarf stars based on (1)
their observed spatial distribution (similar to old disk stars), 
luminosity function and mass distribution;
(2) H and He white dwarf atmosphere models (Bergeron {\em et al.} 1995)
that transfer white dwarf properties to the SDSS magnitude system.
We show that white dwarfs outnumber quasars for $r' \lesssim 17$ at
moderate Galactic latitude. 
Hot white dwarfs can be easily picked out (and separated from
low redshift quasars) from their positions in color space.

 We simulate the distribution of compact emission line galaxies
based on (1) the luminosity and  emission line strength distribution  
from SSG; (2) the continuum slope distribution from McQuade {\em et al.} (1995);
and (3) the size distribution of the UM sample (Salzer {\em et al} 1989).
We show that down to the SDSS spectroscopic survey limit of $i'\sim 19$,
about half of the emission line galaxies will be so compact that 
they cannot be distinguished from stars based on SDSS images.
Those galaxies occupy a similar place in color space to that of  quasars, 
but the majority
of them are redder than low-redshift quasars.

 We simulate the distribution of quasars based on (1) a quasar
spectral model, which includes the observed distributions of 
power-law continuum, optical/UV emission lines and large number
of absorption systems due to intervening neutral hydrogen;
and (2) the evolution of the luminosity function of quasars.
We show how the colors of quasars evolve with redshift.
The low redshift ($z < 2.2$) quasars can be easily separated
from stars and hot white dwarfs, when more than two colors are
used in the selection.
The colors of high-redshift quasars are dominated by  
absorption blueward of Ly $\alpha$ emission, and are well-separated
from the stellar locus.
The colors of quasars with $2.5 < z < 3.0$, however,
are easily confused with hot halo stars, especially halo blue horizontal
branch stars, and relative cool ($T_{eff} \sim 7000$ K ) white dwarfs.
We also discuss the separation of quasars from other types of
objects using color in more detail,
and suggest that a color cut in $g'-r'$ will eliminate most
of the emission line galaxies and hot white dwarfs, which 
themselves are also well-separated from the stellar locus, and
could be selected as quasar candidates otherwise. 
For $2.5 < z < 3.0$, on the other hand, quasars fall 
very close to the stellar locus. 
This effect is also a function of Galactic position, with 
stellar population  depending on direction.
 The simulation is used to establish and
test the quasar selection criteria for the SDSS survey.

 We show the relative distribution of different type of objects
in color space.
All four kinds of stellar objects
are distributed basically on the same plane in color space,
as previously pointed out by Newberg {\em et al.} (1999) from
test observations using SDSS filters.
This is due to the fact that (1) the red part of optical stellar spectra
does not deviate much from a black-body; and (2) the power law
continuum that dominates the colors of quasars and CELGs also lies
close to this plane.
Changes in  stellar metallicity, gravity and reddening mostly
result in scatter of the stellar distribution {\em within} this plane.

 The simulation developed in this paper will be thoroughly tested and
improved when comparing with the soon-available SDSS survey data.
It will be a very useful tool both for the study of Galactic structure
and the selection of peculiar objects from the SDSS survey.

 We thank Jim Gunn, Zeljko Ivezic, Robert Lupton, Jill Knapp, Allyn Smith,
Heidi Newberg, Yuen Ng, Jeff Pier, Neil Reid, Gordon Richards, Don Schneider, 
Michael Strauss and Simone Zaggia for very helpful discussions.
We thank Pierre Bergeron for providing theoretical SDSS colors
of white dwarf models; Heidi Newberg and Brian Yanny for their
stellar locus fitting code;
Dawn Lenz for the synthetic colors of
observed white dwarfs; Kevin Krisciunas for sending his catalog in
digital form; Irwin Horowitz for sending  his data in digital form,
and Mike Blanton for providing his 3-D plotting software.
We thank Yuen Ng for his help on stellar simulations.
We especially thank Michael Strauss and Jill Knapp for carefully reading
and commenting an earlier manuscript.
This research is supported in part by NSF grant AST96-16901, and the
University Research Board of Princeton University.

\newpage
\appendix
\section{SDSS Photometric System and Photometric Errors}

The SDSS system comprises five bands ($u', g', r', i', z'$)
that cover the entire range from the atmospheric cutoff in
the ultraviolet to the sensitivity limit of silicon in the
near IR (F96, for a description of the SDSS photometric camera, 
see Gunn {\em et al.} 1998). 
The sensitivity curves of the five bands are shown
in Figure 1. We plot  the total system throughput
including the quantum efficiency of the CCD, the throughput
of the telescope and camera optics,
and the average atmospheric extinction at an airmass of 1.2 for
the altitude of Apache Point Observatory (2800 m).
We also show the low resolution spectra (in the units
of $f_{\lambda}$, arbitrary  normalization) of a quasar with
redshift of 2.8 and a normal F5V type star (for the star/quasar
separation, see \S 6). 

The zeropoint of the SDSS photometric system is based on the
$\rm AB_{95}$ system defined in F96 (see also Oke \& Gunn 1983).
The magnitude is defined by :

\begin{equation}
m = -2.5 \log \frac{\int d(\log \nu) f_{\nu} S_{\nu}}
		{\int d(\log \nu)  S_{\nu}} - 48.60
\end{equation}

\noindent
where $S_{\nu}$ is the system throughput.
They are based on the observations of four F-subdwarf spectrophotometric
standards of Oke \& Gunn (1983). 
Their synthetic SDSS magnitudes are listed in F96.
The SDSS photometric system will be defined by observations with
the SDSS ``Photometric Telescope'', a 50 cm reflector located
at Apache Point Observatory,  during the commissioning phase of the
survey.

The basic properties of each band are listed in Table 2.
The effective wavelength $\rm \lambda_{eff}$ listed in Table 2 is
defined by Schneider {\em et al.} (1983):

\begin{equation}
\lambda_{eff} = \exp \frac{\int d(\ln \nu) S_{\nu} \ln \lambda}
                          {\int d(\ln \nu) S_{\nu}}
\end{equation}

\noindent
which is in some sense halfway between an effective wavelength and
an effective frequency (see also F96).

We need to simulate the photometric errors as observed by
the real survey in order to compare the simulation with the
observations. The photometric errors have two contributions: 
photon noise (from the star, sky, dark current and read noise)
and calibration error (how well the magnitudes are tied to
the primary standards, F96). For an object with  a noiseless
magnitude in band $i$ as $m^{real}_{i}$, we apply the following
procedures to get its observed magnitude  $m^{obs}_{i}$ and
associated photometric error $\sigma^{obs}_{i}$:

1. The total mean number of electrons detected will be :
\begin{equation}
\langle S_{i} \rangle = 1.96 \times 10^{11}\times t \times  Q_{i} \times 10^{-0.4m^{real}_{i}},
\end{equation}
where $t$ is the total exposure time (55 seconds);
$Q_{i}$ is the flux sensitivity quantity defined in F96 (see Table 2):

\begin{equation}
Q_{i}= \int d(\ln \nu) T^{i}_{\nu},
\end{equation}

and $T^{i}_{\nu}$ is the total system throughput including CCD, telescope
and atmosphere.

2. The number of  observed electrons $S_{i}$   
is drawn from a Poisson distribution
$P[\langle S_{i} \rangle]$.

3. The total photon noise will be :
\begin{equation}
N_{i} = [ S_{i} + RN^{2} + n_{eff} \times t \times (sky+dark)_{i}]^{1/2}
\end{equation}
where $RN$ is the read noise of the CCD (close to 7 $e^{-}$, 
Gunn {\em et al.} 1998); 
$t \times (sky+dark)_{i}$  is the total sky and dark current counts
per pixel; and $n_{eff}$ is the effective number of pixels used
 for  PSF photometry (for the normal seeing condition in
APO, $n_{eff} = 28$ pixel, 1 pixel = $0.4''$).

4. The magnitude and error without calibration error are:
\begin{equation}
m^{nocal}_{i} = -2.5 \log [ S_{i} / (1.96 \times 10^{11} \times  t \times Q_{i})]
\end{equation}
\begin{equation}
\sigma^{nocal}_{i} = 1.086 N_{i}/ S_{i}
\end{equation}

5. We assume a random calibration error  $\sigma^{cal}_{i}$= 0.02 mag for all bands.
A random $\Delta(m)$ from a Gaussian distribution is added
to $m^{real}_{i}$: 
\begin{equation}
m^{obs}_{i} = m^{nocal}_{i} + \Delta(m)
\end{equation}
\begin{equation}
\sigma^{obs}_{i} = [(\sigma^{nocal}_{i})^{2} + (\sigma^{cal}_{i})^{2}]^{1/2}
\end{equation}

\section{Simulation of HI Absorption Spectrum of High-redshift Quasar}

We create the synthetic spectrum of absorptions by intervening HI
absorbers along the line of sight to high-redshift quasars, in the
spectral range blueward of $\rm Ly\alpha$ emission.
The absorption systems we consider are $\rm Ly\alpha$ forest systems,
Lyman Limit Systems (LLSs) and damped $\rm Ly\alpha$ systems.
We simulate both the line absorption (up to $\rm Ly_{10}$) and
continuum absorption by those systems. 
The simulation follows procedures similar to those outlined by M\o ller \& Jakobsen (1990),
Fan \& Chen (1993) and Warren {\em et al.} (1994).

A HI absorber is characterized by three parameters: its HI column density
$N_{HI}$, redshift $z$ and Doppler width $b$. 
The number density of the absorption systems evolves with redshift.
The evolution is usually fitted by 
a power law:

\begin{equation}
N(z) = N_{0} (1+z)^{\gamma}
\end{equation}

And the HI column density distribution is also usually expressed as a power law:

\begin{equation}
f(N_{HI})  \propto  N_{HI}^{-\beta}
\end{equation}

The redshift evolution of the $\rm Ly\alpha$ forest system is well established.
On the other hand, the results on the evolution of LLS and damped system 
remain controversial.
Sargent {\em et al.} (1989) found $\gamma=0.68$, 
consistent with  no evolution in co-moving density for LLS, 
while Lanzetta (1991) found strong evolution for $z>2.5$ ($\gamma=3.5$). 
Storrie-Lombardi {\em et al.} (1994) found a evolution rate 
somewhere in between ($\gamma=1.55$).
The evolution rate of LLS is important for the quasar colors, since
the optically thick continuum absorption has the biggest effect on
the colors of high-redshift quasars.
Storrie-Lombardi {\em et al.} (1994) used the APM sample, 
which includes a large number of quasars at $z>4$, so we will use their
statistics on $\gamma$ and $\beta$ for the LLS and damped systems.
The statistics we adopt in the simulation of HI absorbers are
summarized in Table 3.

The absorption spectrum is simulated as follows:

1. For a quasar with emission redshift $z_{em}$, the total number $N$ of  each kind of
absorber along the line of sight is calculated by drawing randomly from
a Possion distribution $P(N_{ave})$, where $N_{ave}$ is the average number expected:
$N_{ave}=\int_{0}^{z_{em}} N(z) dz$.

2. For each of the $N$ absorbers, its redshift and column density are randomly selected
from the distribution according to (B1) and (B2), respectively.

3. For each absorber, the opacity profile $\tau_{\lambda}$ of the first ten lines in the Lyman
series are calculated. A Voigt profile with width $b$ and natural broadening
are assumed. The Lyman continuum absorption of each line under the Lyman Limit
is also calculated. Then, the line and continuum absorptions are redshifted,
with the contribution on opacity from all of the absorbers  co-added, and
finally the absorption is applied to the emission line + continuum
spectrum of the quasar.
In this process, two important parameters, $D_{A}$ and $D_{B}$, the 
continuum depression between $\rm Ly\alpha$ and $\rm Ly\beta$, and
between $\rm Ly\beta$ and Lyman Limit, are also calculated:

\begin{equation}
D_{i} = \langle 1-e^{-\tau_{\lambda}}\rangle_{\lambda}, \hspace{1.5cm}
\left\{ \begin{array}{ll}
		1050 \rm \AA < \lambda < 1170 \rm \AA,  & i=\rm A; \\
		920 \rm \AA < \lambda < 1015 \rm \AA, & i=\rm B \\
	\end{array}
\right.
\end{equation}

$D_{A}$ and $D_{B}$ provide a good check on how well we are simulating
the overall level of absorption. Figure 22 plots our simulated distributions
of $D_{A}$ and $D_{B}$ against observations. Both the average and scatter
of the parameters agree very well between simulations and observations.
The simulation thus can reproduce the effect on quasar colors and their scatter
caused by the intervening absorptions.

\newpage

\begin{center}
Table 1. Stellar Population Parameters

\begin{tabular}{cccccc}\\ \hline \hline
Pop. & age & SFR & Z & density profile & normalization at $b=90^{\circ}$ \\
     & Gyr &     &   &                 & for $\rm M > 0.15 M_{\odot}$ \\
\hline
halo & 16 -- 10 &  const. & 0.0004--0.003& de Vaucouleurs $q$=0.8 & 14000  star/deg$^2$ \\
disk & 10 -- 0.5&  const. & 0.007 -- 0.03& exponential, $h$ = 3.5 kpc & 3300 star/deg$^2$  \\
      &          &         &              & $z_{0} =95(1 + t /0.5\ \rm Gyr )^{2/3}$ pc &  \\ \hline \hline
\end{tabular}

\end{center}

\begin{center}
Table 2. Filter Properties

\begin{tabular}{cccccc} \\ \hline \hline
Filter  & $u'$  & $g'$  & $r'$ &  $i'$ &  $z'$ \\ \hline
$\lambda_{eff}$ (\AA) & 3543 & 4770 & 6231 & 7625 & 9134 \\
FWHM (\AA)      & 567  & 1387 & 1373 & 1526 & 950  \\
$Q$             & 0.0185 & 0.118 & 0.117 & 0.0874 & 0.0223 \\
$t \times (sky+dark)$/pixel & 45 & 401 & 690 & 1190 & 1120 \\
$m_{lim} (5 \sigma)$ & 22.3 & 23.3 & 23.1 & 22.7 & 20.8 \\
$A_{i}/A_{V}$ & 1.593 & 1.199 & 0.858 & 0.639 & 0.459 \\ \hline \hline
\end{tabular}
\end{center} 

\begin{center}
Table 3. Properties of HI Absorbers

\begin{tabular}{ccccccc}\\ \hline \hline
         & $\log N_{HI}$ (cm$^-2$) & $ N_{0}$ & $\gamma$ & $\beta$ & $b$ (km s$^{-1})$ & Ref \\ \hline
 $\rm Ly\alpha$ forest & 13 -- 17.3 & 50.3 & 2.3 & 1.41 & 30 &  M\o ller \&  Jacobsen (
1990) \\
 LLS & 17.3 -- 20.5 & 0.27 & 1.55 & 1.25 & 70 & Storrie-Lombardi {\em et al.} (1994)
\\
damped  $\rm Ly\alpha$ system & 20.5 - 22 & 0.04 & 1.3 & 1.48 & 70 & Storrie-Lombardi
 {\em et al.} (1994) \\ \hline \hline
\end{tabular}

\end{center}

\begin{figure}
\vspace{-4cm}

\plotone{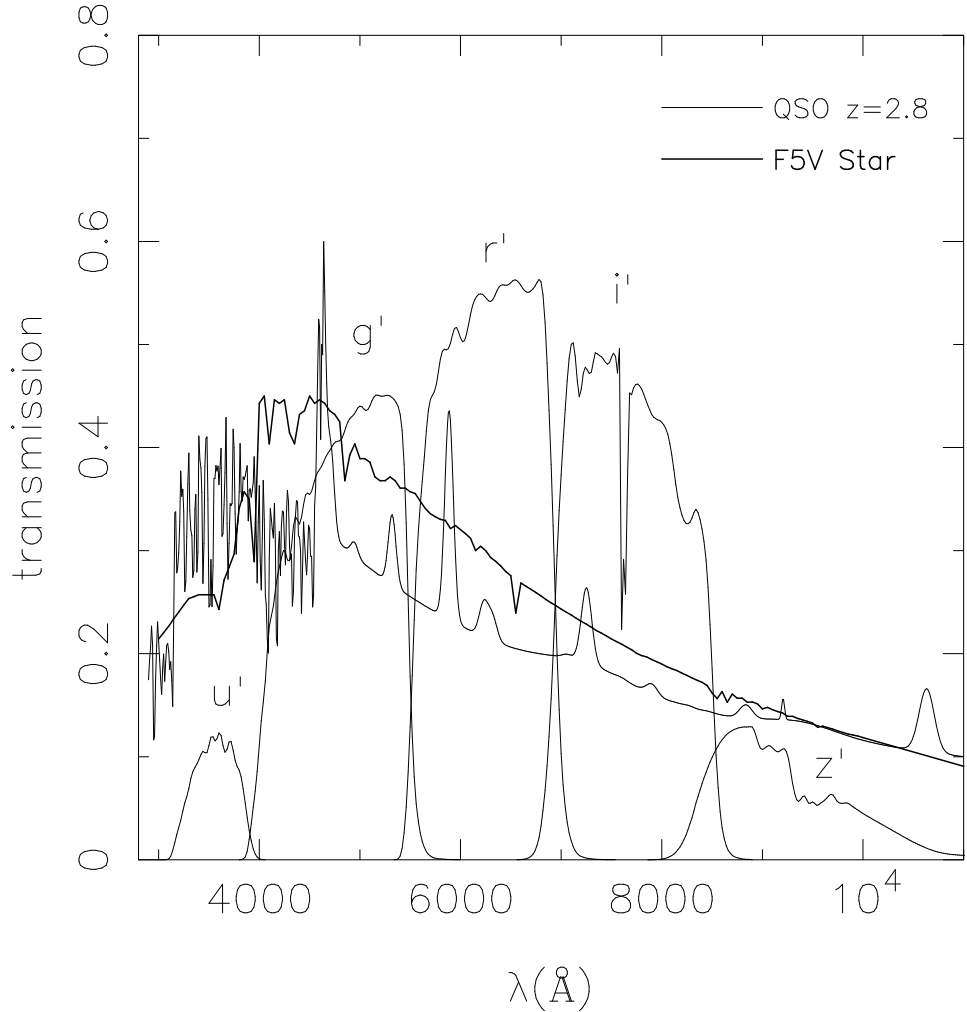}

\vspace{1cm}
\caption{Response function of the SDSS photometric system,
including the telescope efficiency, CCD quantum efficiency,
filter transmission and the atmospheric transmission at
1.2 airmass at the altitude of Apache Point Observatory
(Adapted from Fukugita {\em et al.} 1996). Also plotted
are the spectral energy distribution of a F5V star and
a quasar at $z=2.8$. Note the striking similarity of
the two spectra, which illustrates the difficulty in
separating quasars at that redshift range with stars
from the broad-band colors.}
\end{figure}

\begin{figure}
\vspace{-2cm}

\plotone{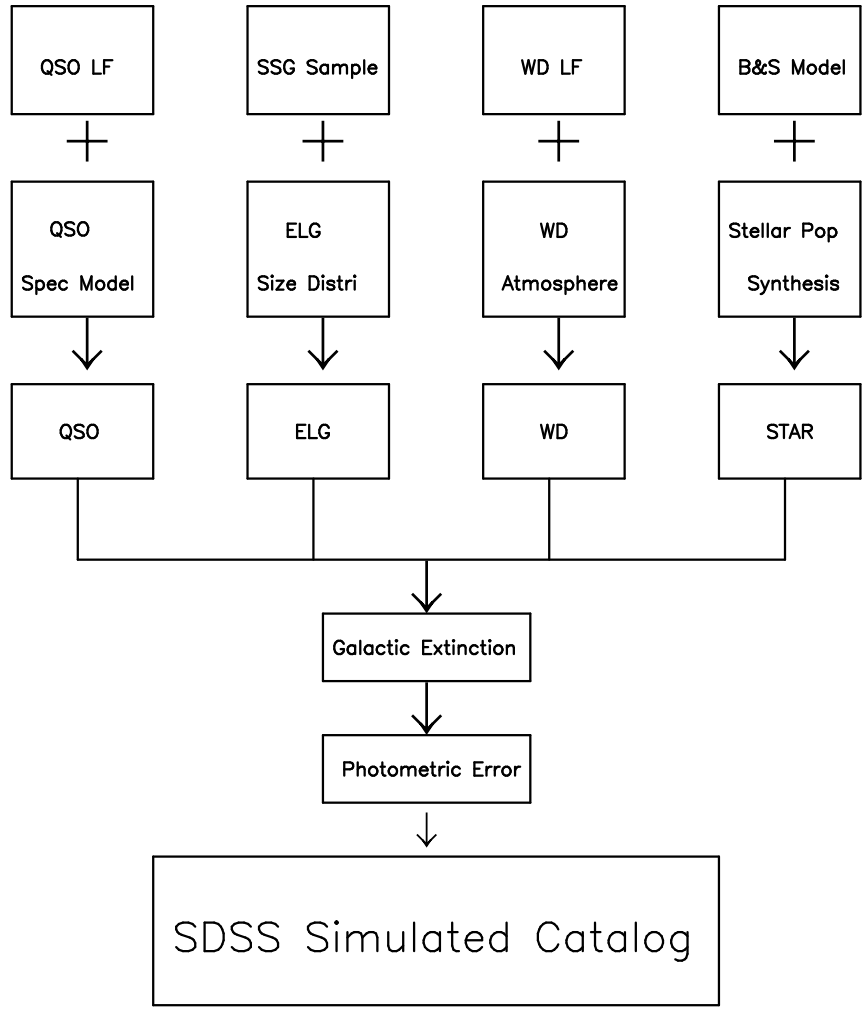}

\vspace{1cm}
\caption{A schematic chart of the structure of the color space simulation programs.}
\end{figure}

\begin{figure}
\vspace{-4cm}

\plotone{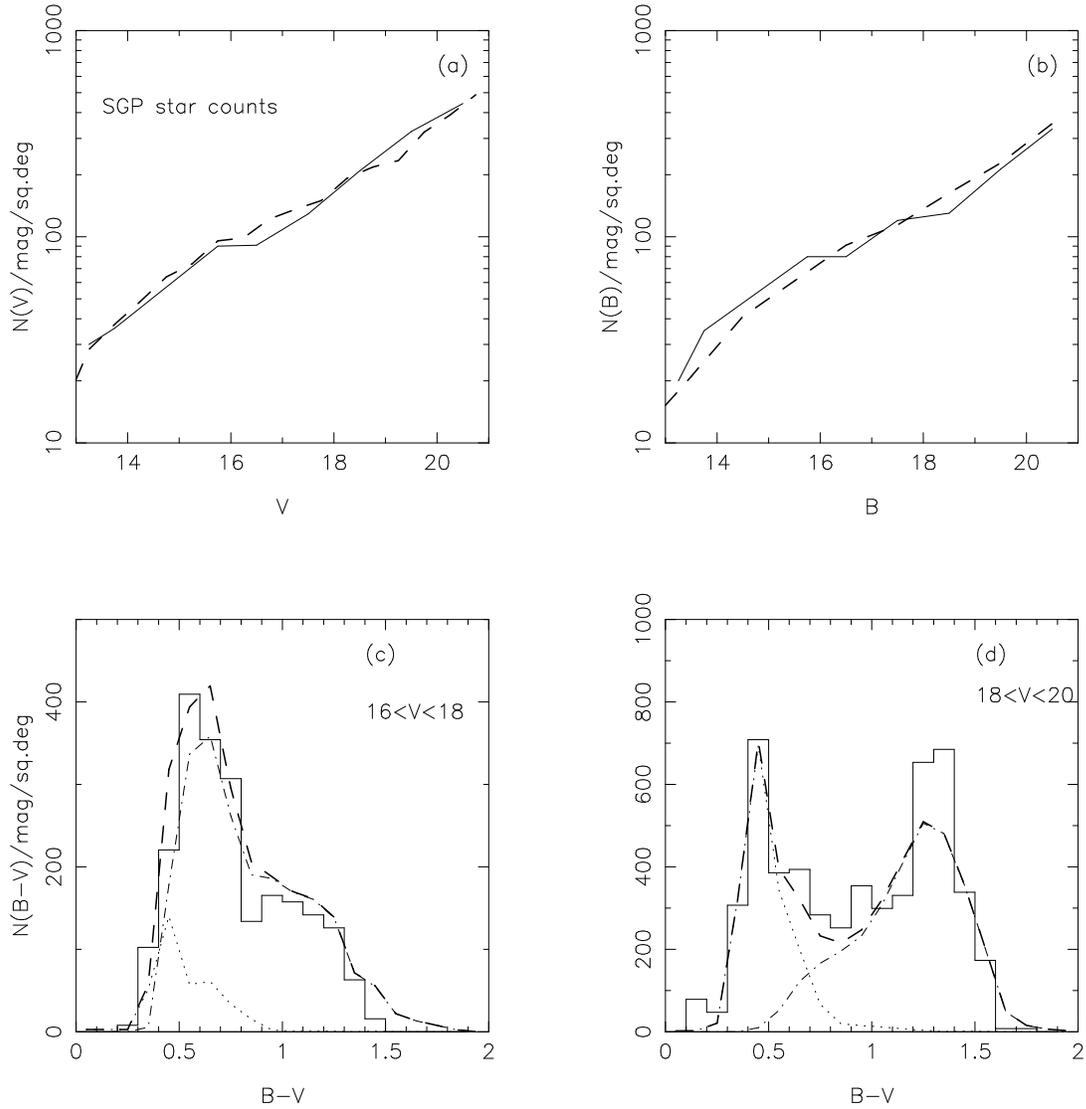}

\vspace{1cm}
\caption{Comparison of the star counts predicted by the simulation with
the observation of ESO Imaging Survey Patch B towards the South Galactic
Pole (Parandoni {\em et al.} 1998). The solid curves are the observed
star counts. The thick dashed lines are the predicted total star counts.
The thin dotted lines  and dash-dotted lines are the contributions from
the halo and disk populations, respectively.}

\end{figure}
 
\begin{figure}
\vspace{-4cm}

\plotone{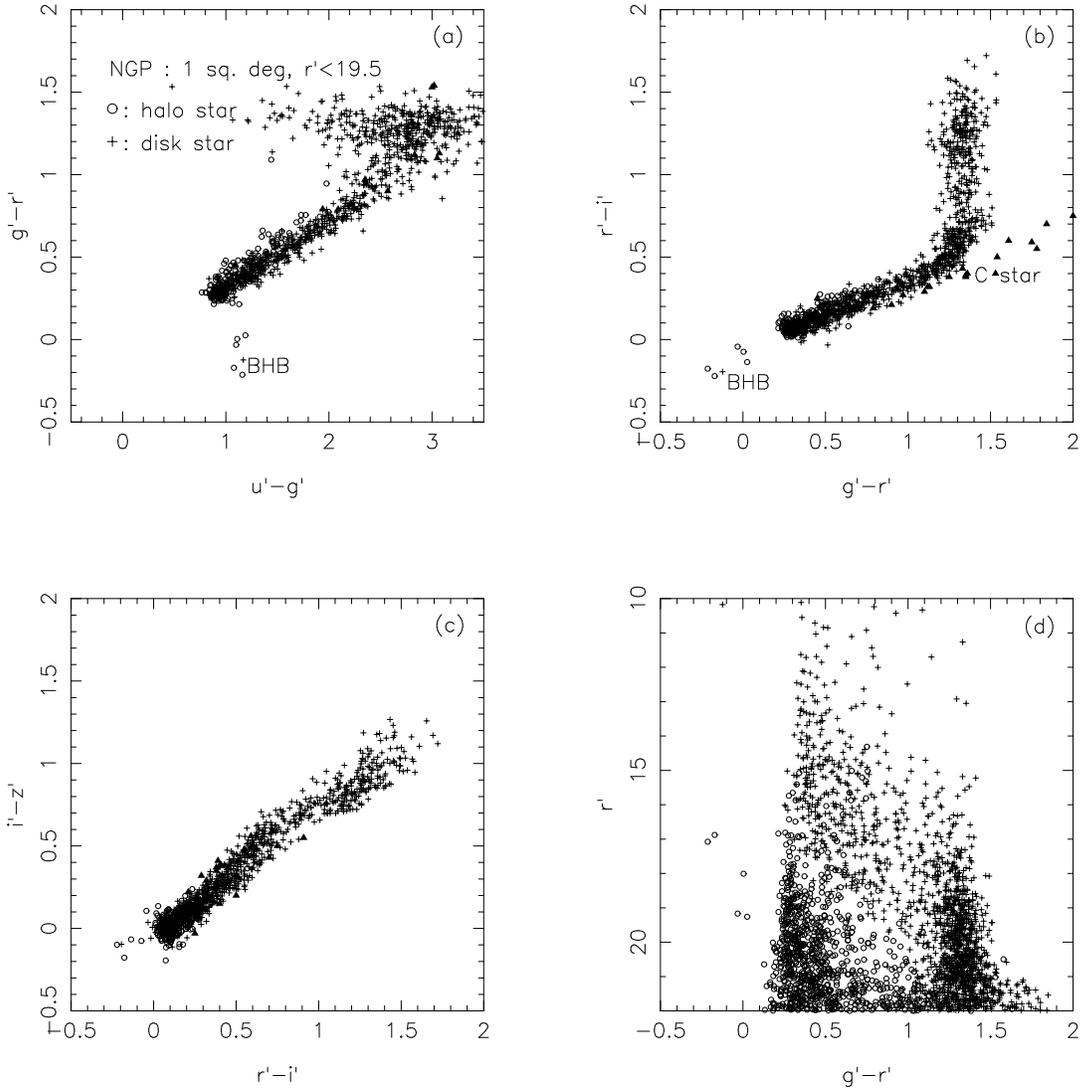}

\vspace{1cm}
\caption{Color-color diagrams and color-magnitude diagram of
a 1 deg$^2$ simulation towards the North Galactic Pole, including
the photometric errors for the SDSS observation and interstellar extinction.
The solid triangles represent colors of Carbon stars observed by Krisciunas
{\em et al.} (1998).}
\end{figure}
\newpage
\begin{figure}
\vspace{-4cm}

\plotone{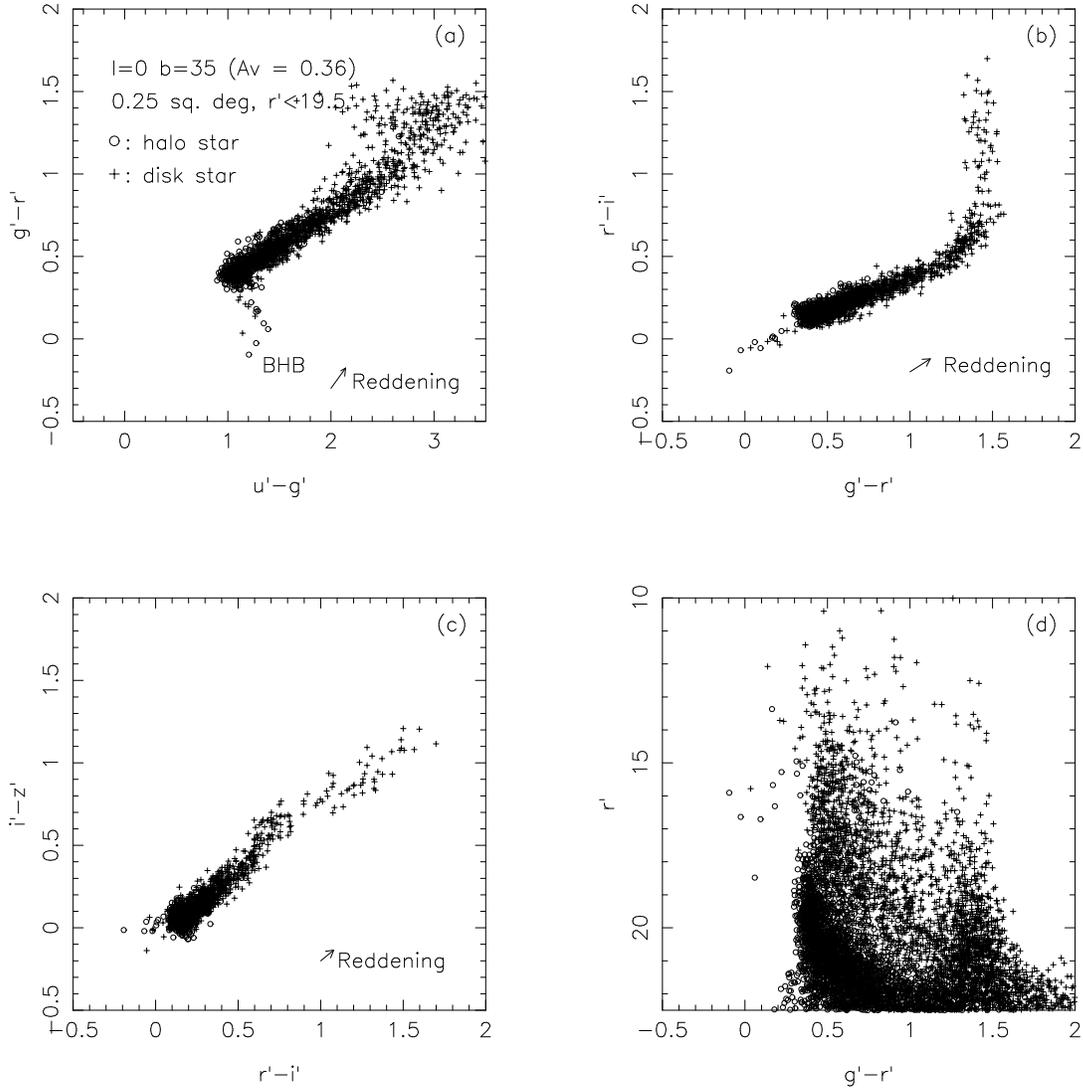}
\vspace{1cm}
\caption{Color-color diagrams and color-magnitude diagram of
a 0.25 deg$^2$ simulation towards $l=0^{\circ}$ and $b=35^{\circ}$.
It has the highest stellar density and extinction ($A_{V}=0.36$) of the whole SDSS
survey region.}
\end{figure}
\newpage
\begin{figure}
\vspace{-4cm}

\plotone{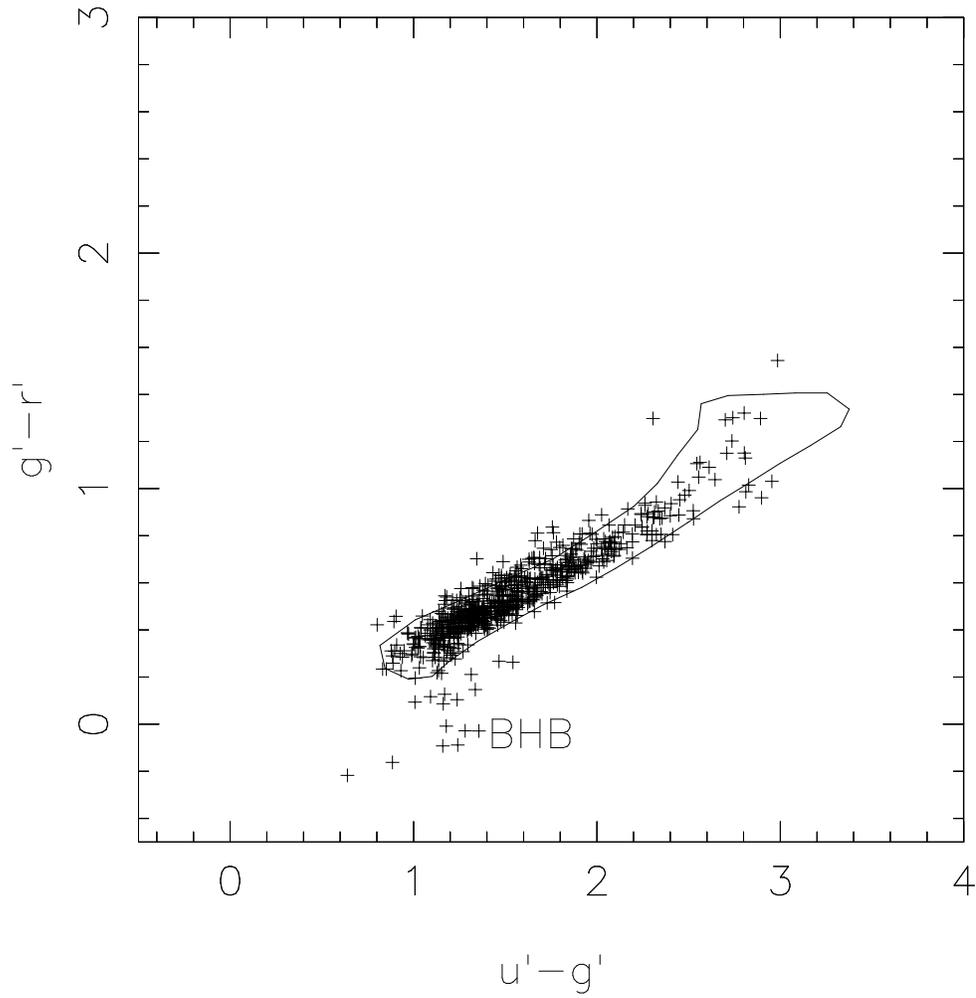}

\vspace{1cm}
\caption{Comparison of the observations (+) of stellar colors by
Krisciunas {\em et al.} (1998) with our simulation.
(the boundary of stellar locus estimated from the simulation in Fig.4).}

\end{figure}
\newpage
\begin{figure}
\vspace{-4cm}

\plotone{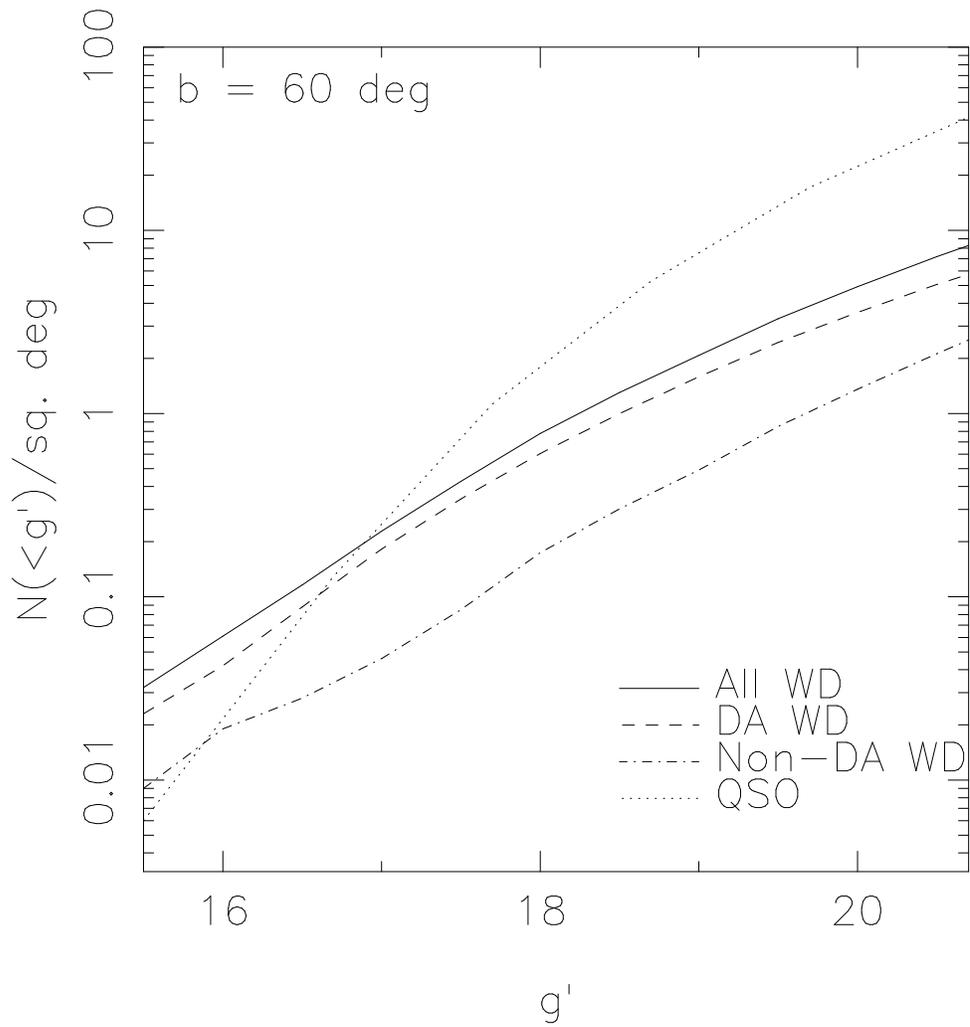}
\vspace{1cm}
\caption{Number-magnitude relation for DA and Non-DA white dwarfs 
at $b=60^{\circ}$. We also compare it with that of quasars. 
White dwarfs overnumber quasars for $g' < 17$.}

\end{figure}
\newpage
\begin{figure}
\vspace{-4cm}

\epsfysize=600pt \epsfbox{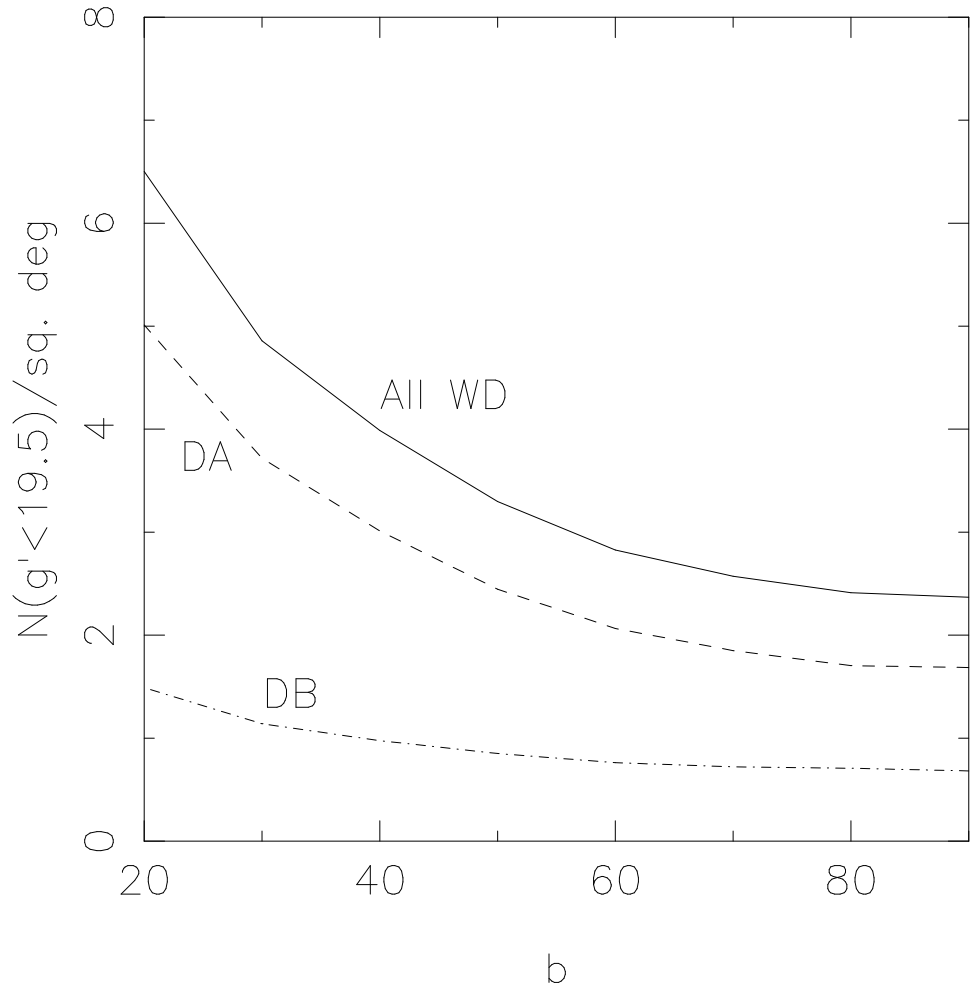}
\vspace{1cm}
\caption{The number density of white dwarfs with $g'<19.5$ as 
a function of Galactic latitude.}

\end{figure}
\newpage
\begin{figure}
\vspace{-4cm}

\epsfysize=600pt \epsfbox{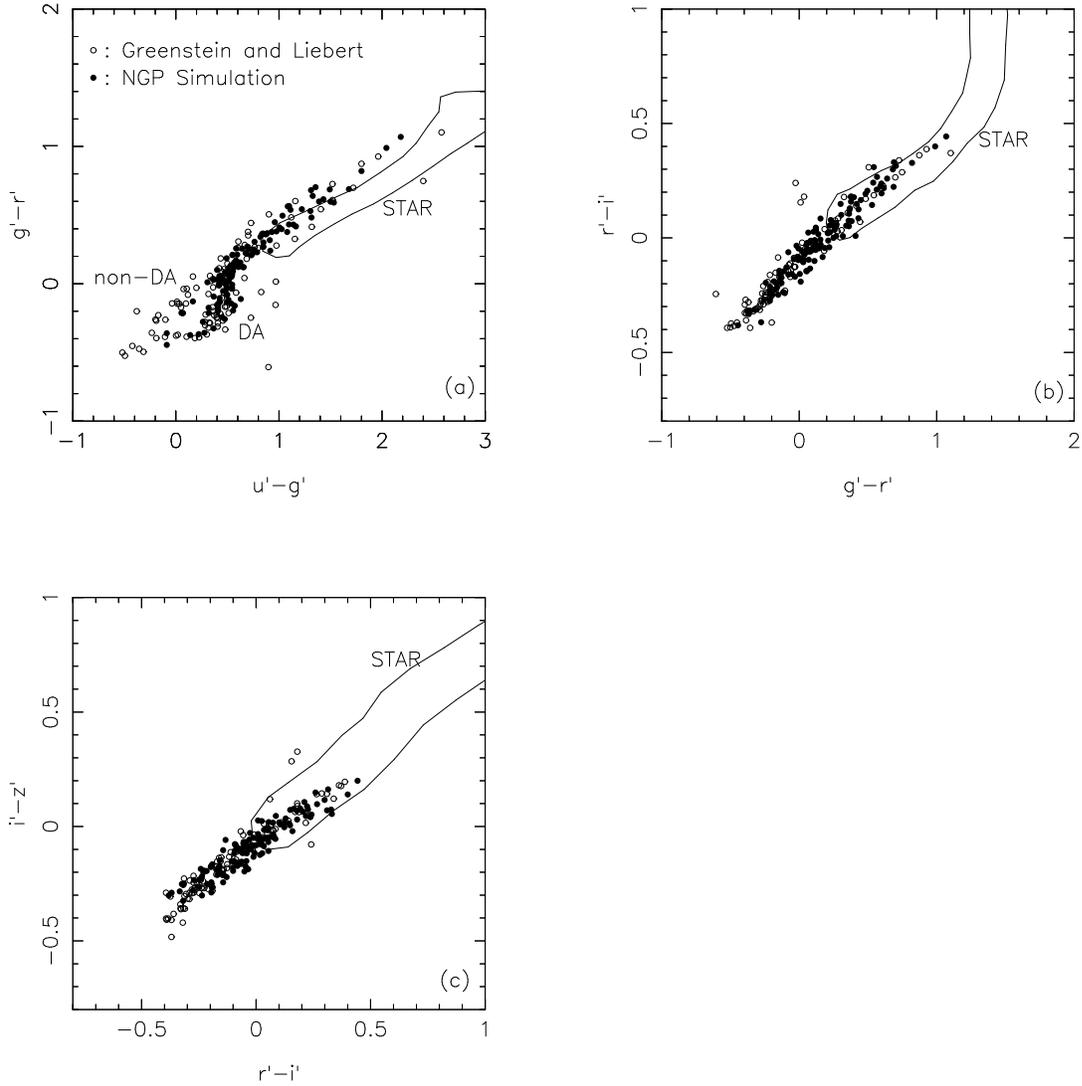}
\vspace{1cm}
\caption{White dwarf distribution on the color-color diagrams in
the SDSS system. Filled circles are from a simulation of 20 deg$^2$
towards $b=60^{\circ}$, with $g'<22.5$;
open circles are the synthetic colors of white dwarfs from
the spectrophotometric atlas of Greenstein \& Liebert (1989),
adapted from Lenz {\em et al.} (1998).
Regions labelled as STAR are the approximate boundaries of stellar
loci (see Figures  4 and 17).}
\end{figure}

\newpage
\begin{figure}
\vspace{-4cm}

\epsfysize=600pt \epsfbox{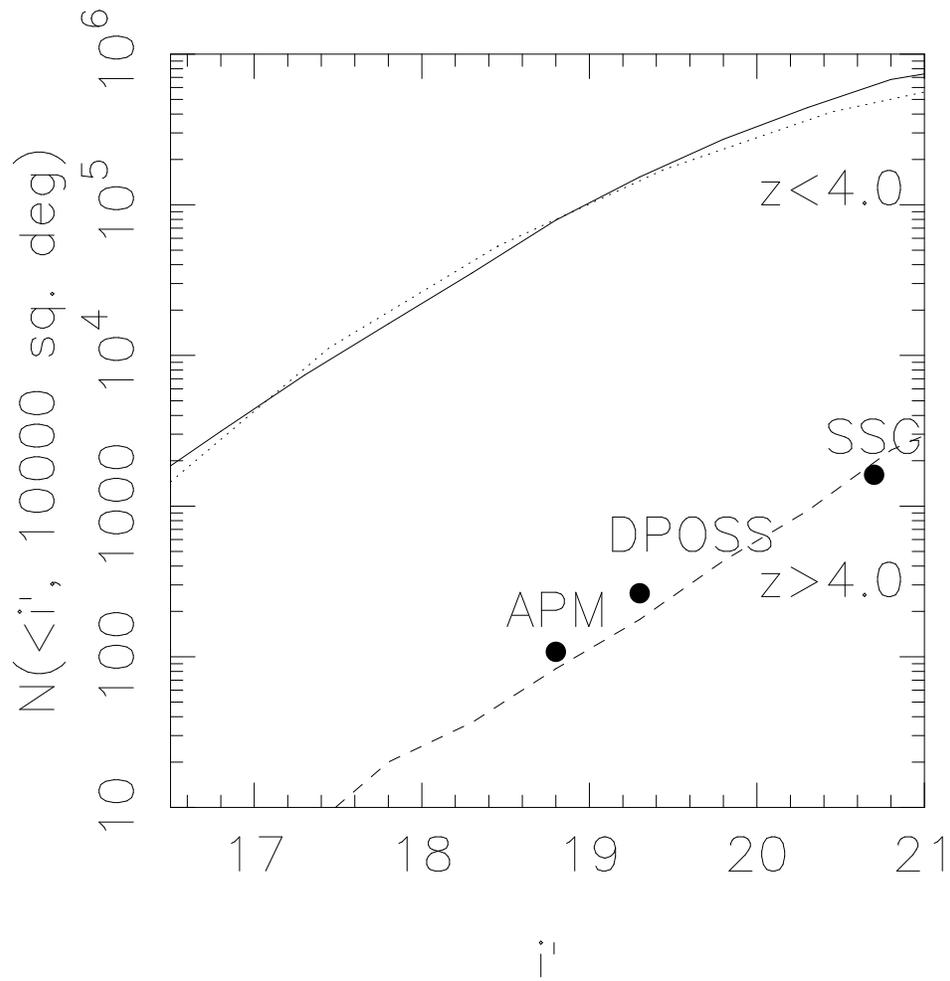}
\vspace{1cm}
\caption{Predicted number-magnitude distributions of low ($z<4$) and
high ($z>4$) redshift quasars.  They are compared with observed
counts: the low-redshift results of Crampton {\em et al.} (1987, dotted line),
and the SSG, DPOSS and APM surveys for $z>4.0$.}

\end{figure}
\newpage
\begin{figure}
\vspace{-4cm}

\epsfysize=550pt \epsfbox{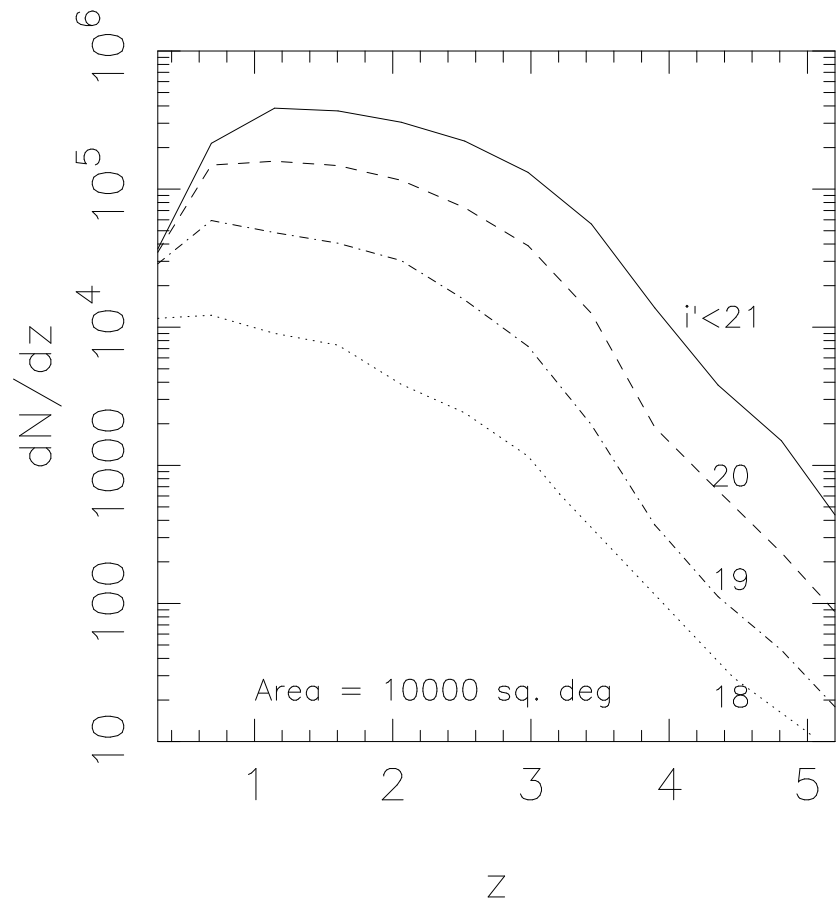}
\vspace{1cm}
\caption{Predicted quasar number counts as a function of redshift for several
different magnitude cuts for the SDSS survey.}

\end{figure}

\newpage

\begin{figure}
\vspace{-4cm}

\epsfysize=600pt \epsfbox{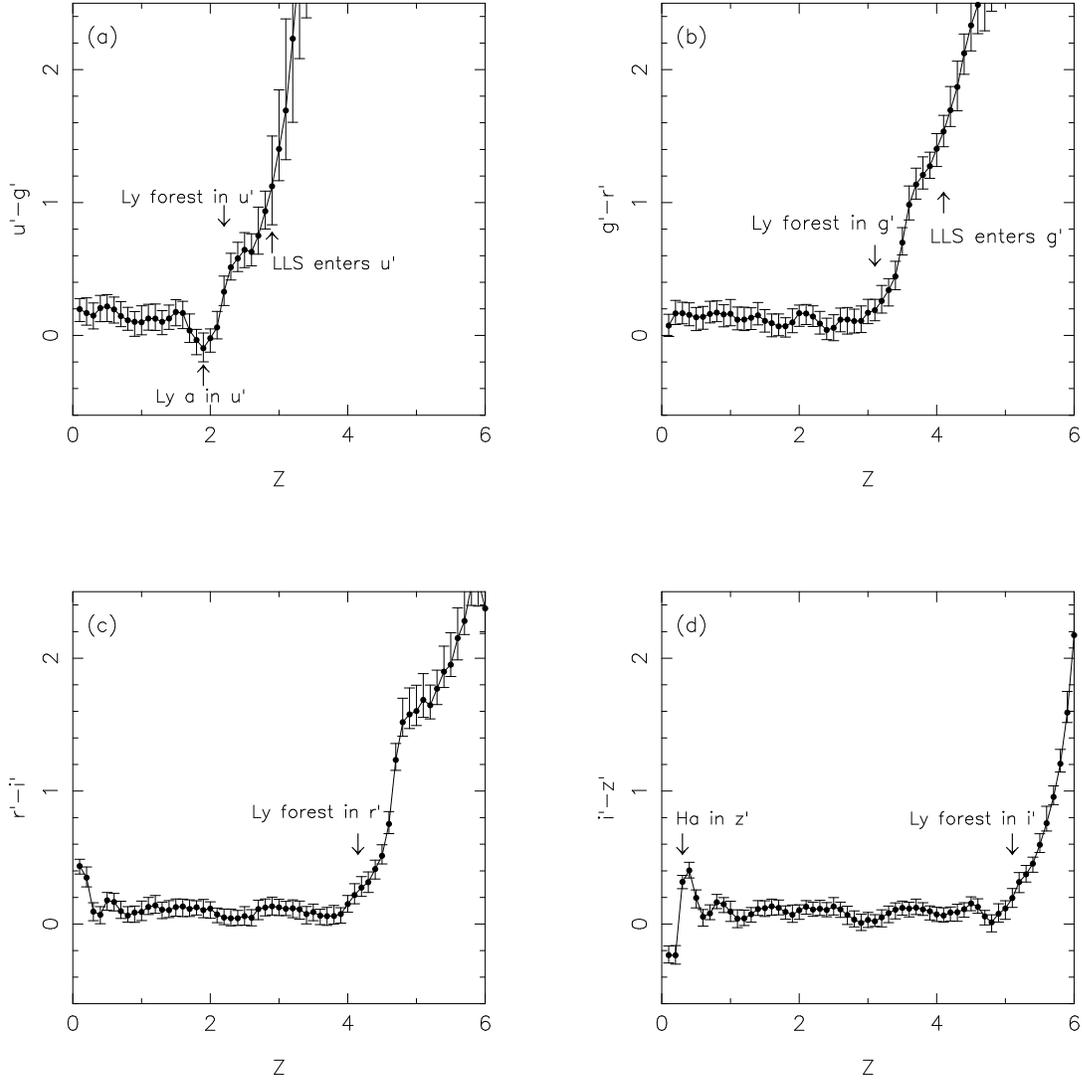}
\vspace{1cm}
\caption{Color evolution of quasars as a function of redshift.
The figures show the median and 68\% scatter 
calculated from simulation of 100 quasars at each redshift.}
\end{figure}

\newpage
\begin{figure}
\vspace{-4cm}

\epsfysize=600pt \epsfbox{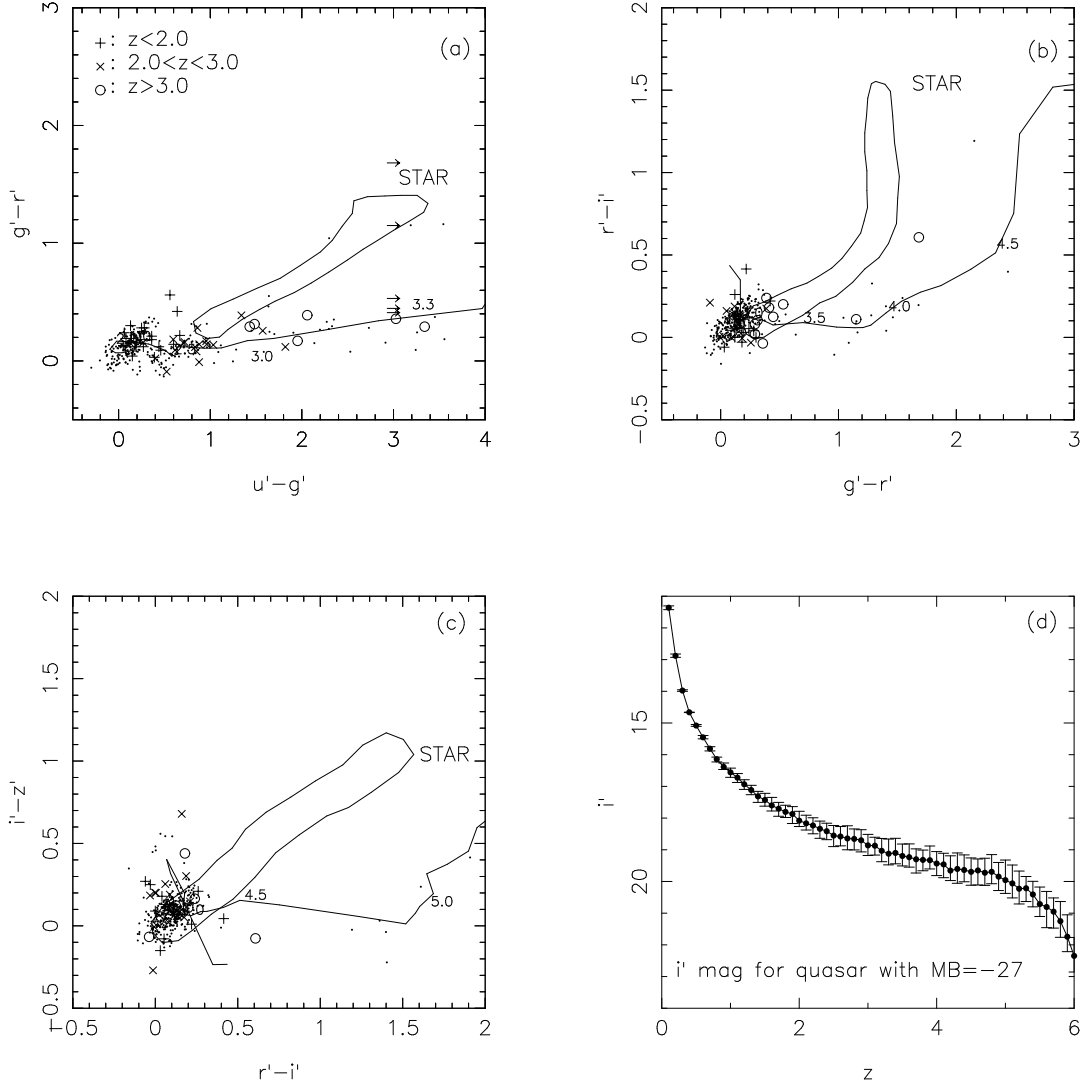}
\vspace{1cm}
\caption{(a)-(c) Color-color diagrams of quasars, from a simulation
of 10 deg$^2$ with $i'<19.5$ (small dots), compared with observations
in SDSS filters of known quasars (bigger symbols) at various redshifts.
Arrows represent the lower limits in $u'-g'$ due to non-detections in
$u'$ observations.
The solid lines labeled with redshift are the median tracks of quasar
colors calculated from Figure 12.
Regions labelled as STAR are the approximate boundaries of stellar
loci.
(d) The median and 68\% scatter of $i'$ magnitude of
quasar with $M_{B}=27$ (similar to 3C273), calculated
from 100 realizations at each redshift. }

\end{figure}
\newpage
\begin{figure}
\vspace{-4cm}

\epsfysize=600pt \epsfbox{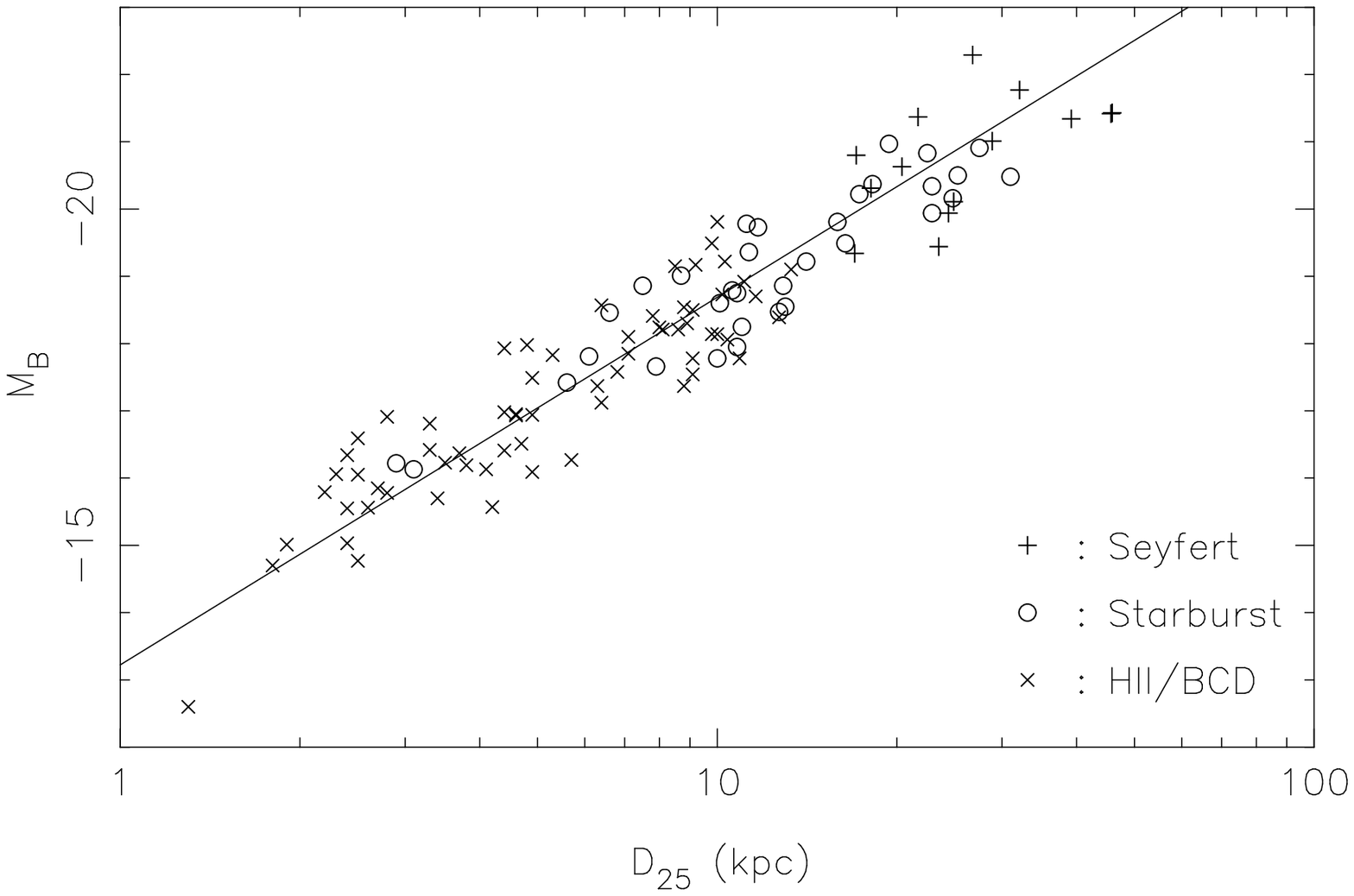}
\vspace{1cm}
\caption{Relation between $M_{B}$ and isophotal diameter $D_{25}$
from the emission line galaxy sample of Salzer {\em et al.} (1989).
Different types of galaxies are indicated by different symbols.
They follow the same best-fit relation (eq.12).}
\end{figure}
\newpage
\begin{figure}
\vspace{-4cm}

\epsfysize=600pt \epsfbox{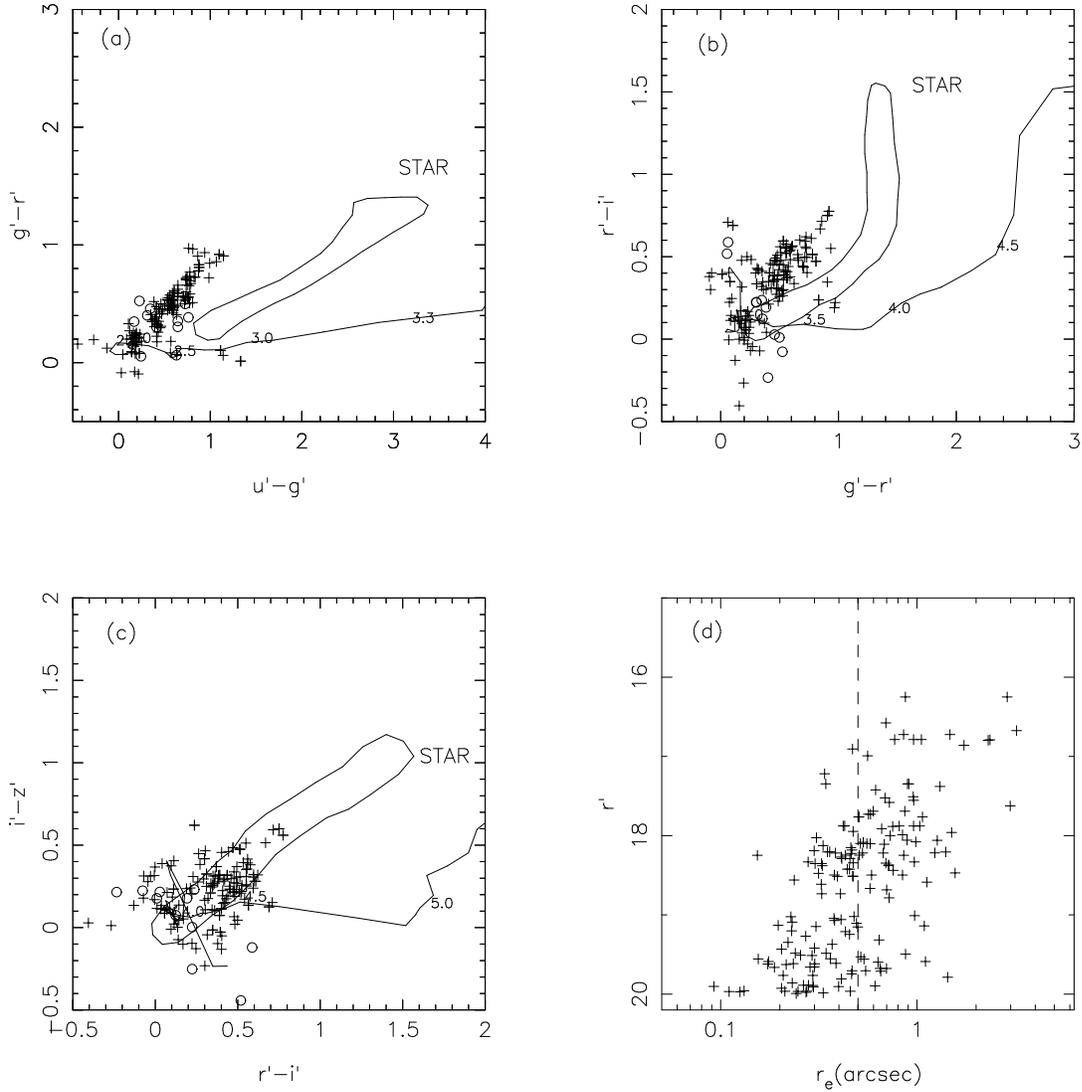}
\vspace{1cm}
\caption{The color-color diagrams and size distribution of
simulated CELGs (+). The observed colors of
CELGs from Hall {\em et al.} (1996, transferred to SDSS bands)
are plotted as open circles.
We also show the approximate stellar loci and the quasar tracks
at different redshift in the diagrams.}
\end{figure}
\newpage
\begin{figure}

\epsfysize=500pt \epsfbox{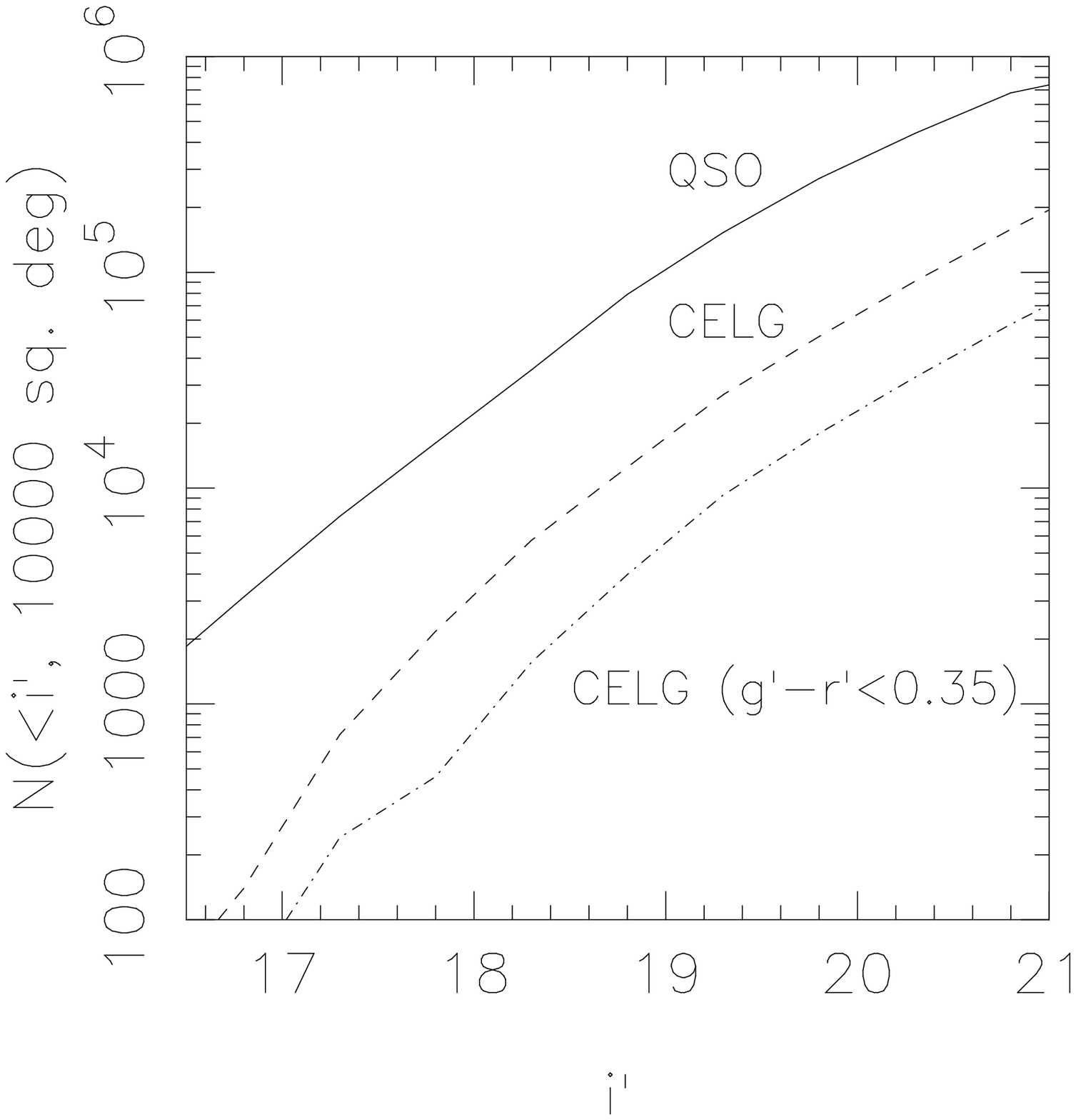}
 \vspace{1cm}
\figcaption{Relation between $M_{B}$ and isophotal diameter $D_{25}$
from the emission line galaxy sample of Salzer {\em et al.} (1989).
Different types of galaxies are indicated by different symbols.
They follow the same best-fit relation (eq.12).}

\end{figure}
\newpage
\begin{figure}

\epsfysize=500pt \epsfbox{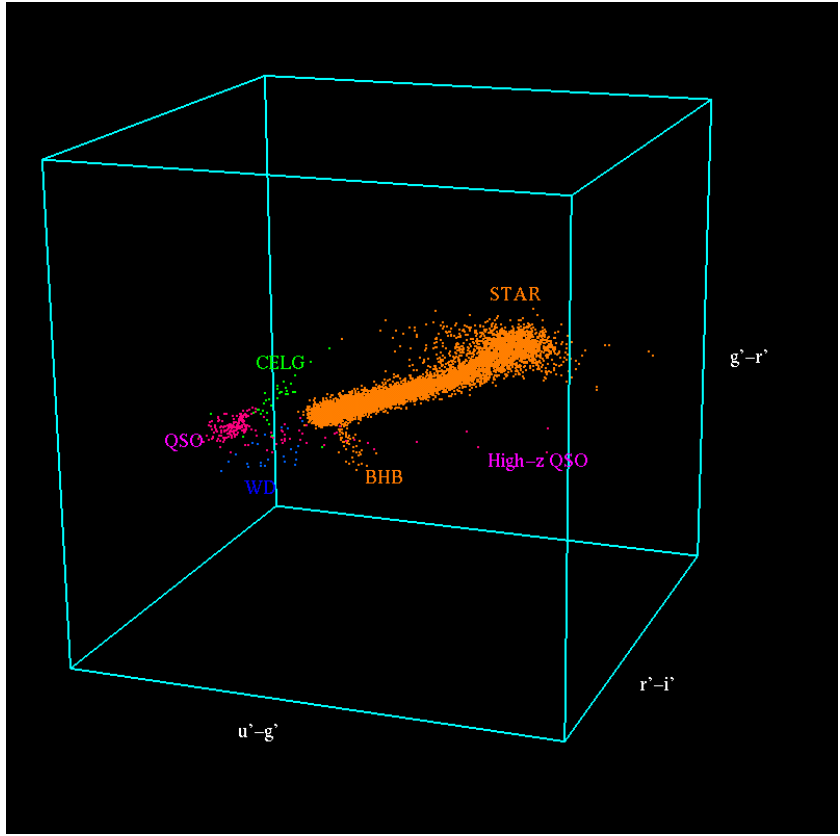}
 \vspace{1cm} 
\caption{Distribution of stellar objects in $u'-g'$, $g'-r'$, $r'-i'$ color space
in a simulation of stars, white dwarfs, quasars, and CELGs for a 10 deg$^2$
region towards the North Galactic Pole. }
\end{figure}
\newpage
\begin{figure}
\vspace{-4cm}

\epsfysize=600pt \epsfbox{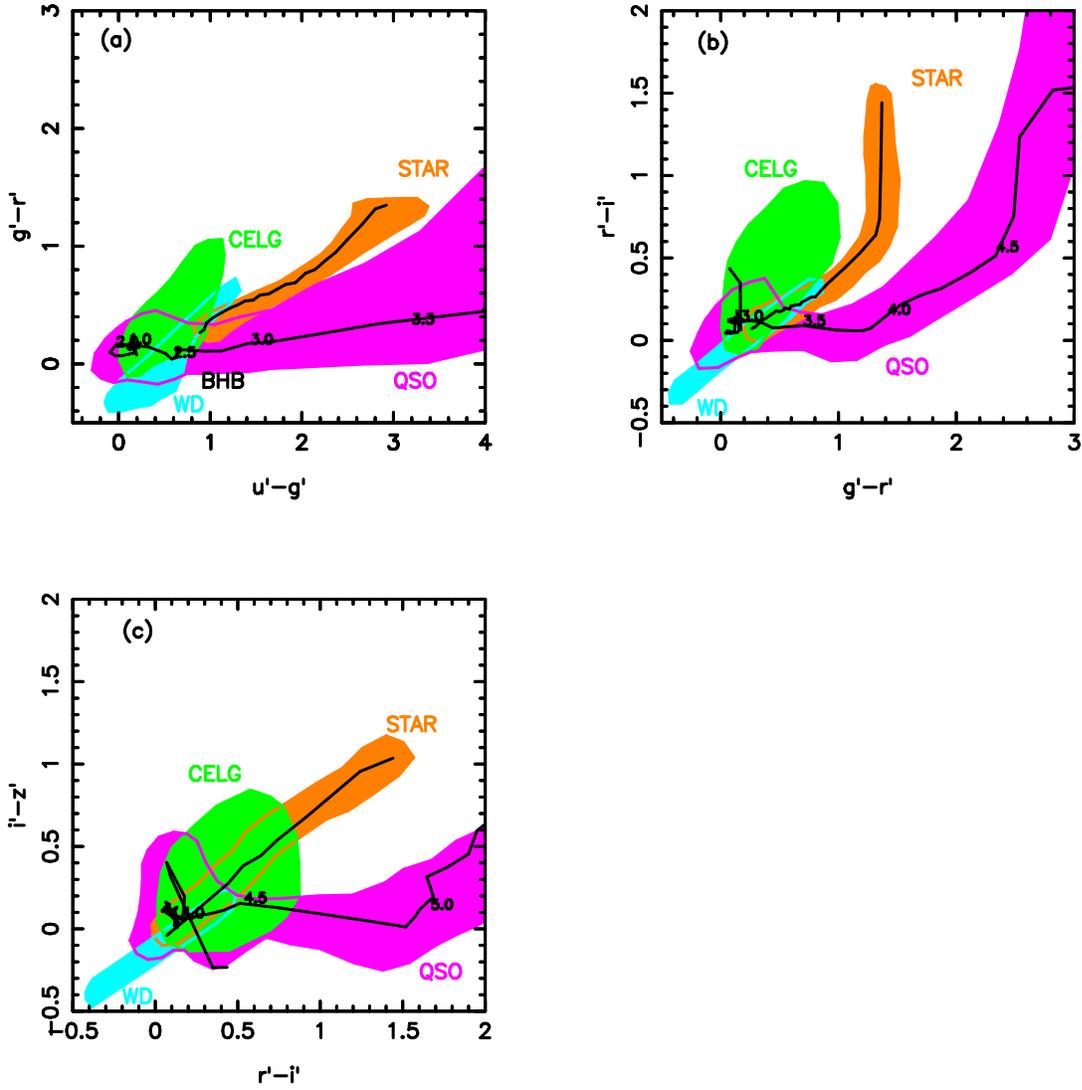}

 \vspace{1cm} 
\caption{ The approximate locus of each kind of objects on the 
color-color diagrams estimated from the simulation shown in Figure 16.
The solid lines going through the stellar loci are the best-fitted
stellar locus points using the algorithm of Newberg \& Yanny (1997).
The lines going through the quasar loci are the median quasar tracks as
a function of redshift.}
\end{figure}

\newpage
\begin{figure}
\vspace{-2cm}

\epsfysize=500pt \epsfbox{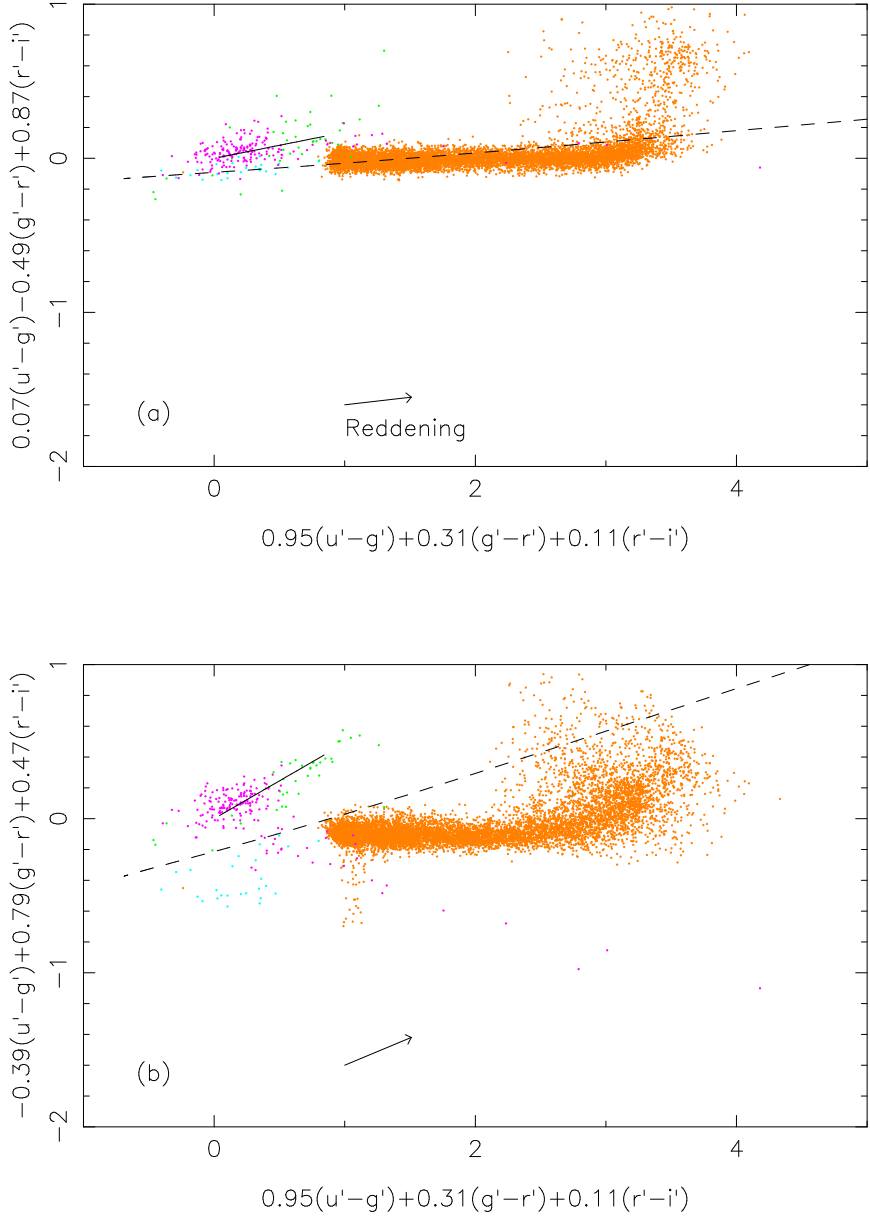}
\vspace{1cm}
\caption{Projections of the stellar objects (from the simulation
shown in Figure 16) onto the $c1-c2$ and
 $c1-c3$ plane (see text). 
Normal stars (orange points), white dwarfs (blue),  CELGs (green) and quasars (magenta)
distributed near to a fundamental plane $c2 \sim 0$ in the
SDSS color space.
The dashed  lines represent the colors of
black bodies for $1000 < T < 30000$ K; and solid lines represent
the color of a power-law spectra with $0 < \alpha < 2$.}

\end{figure}
\newpage
\begin{figure}
\vspace{-4cm}

\epsfysize=600pt \epsfbox{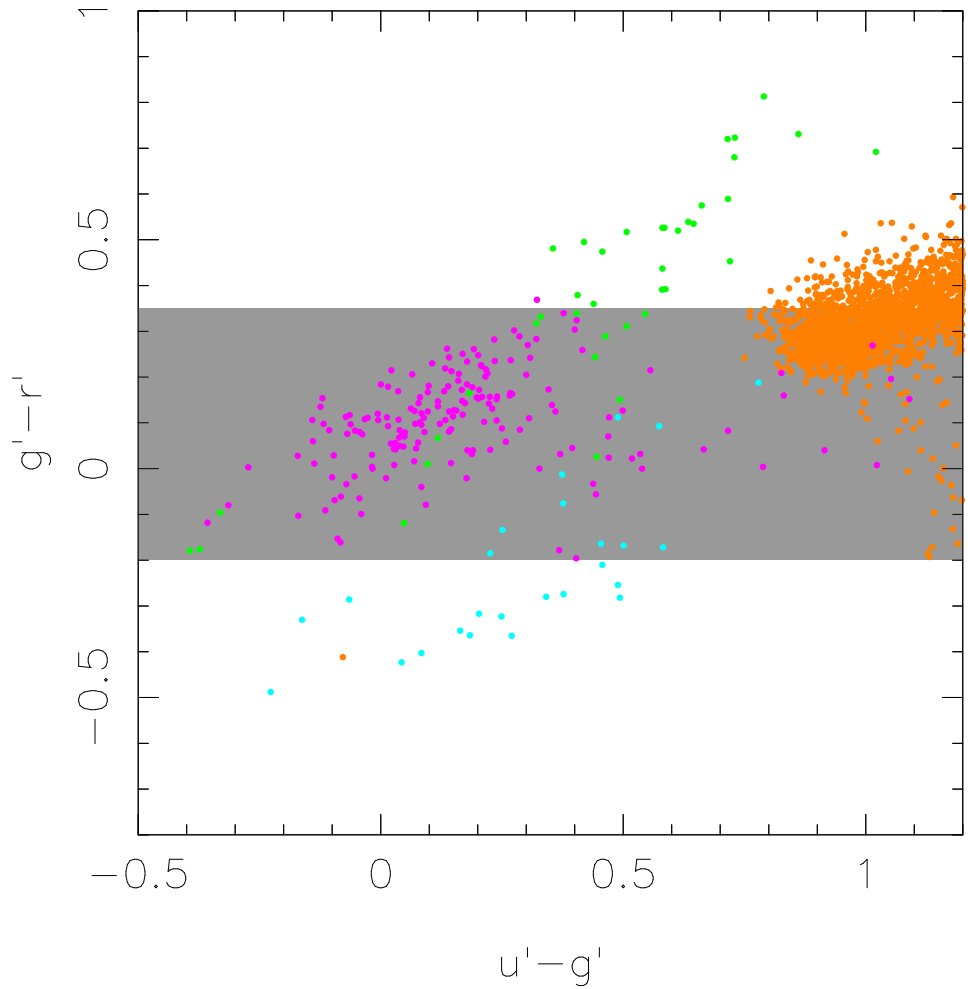}
\vspace{1cm}
\caption{The region in the color-color diagram where the 
white dwarf (blue), star (orange), CELG (green)  and quasar (magenta) loci intersect.
Using a color cut $-0.20 < g'-r' < 0.35$, more than 50\% of
CELGs and white dwarfs are eliminated, but very few quasars
are dropped at the same time.}
\end{figure}
\newpage

\begin{figure}
\vspace{-4cm}

\epsfysize=600pt \epsfbox{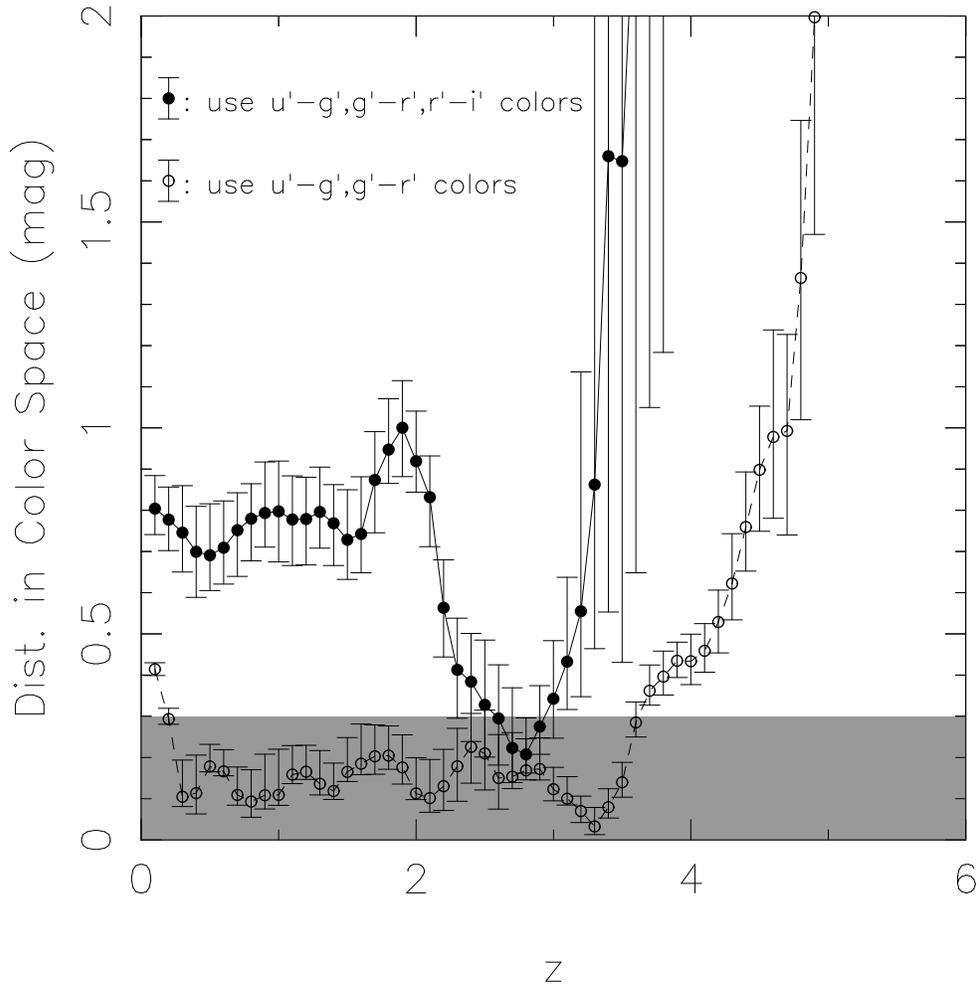}
\vspace{1cm}
\caption{ The distances from the quasar locus to the stellar locus
based on three (filled circles) and two (open circles) 
SDSS colors.
Quasars overlap with stellar locus in the redshift range
$2.5 < z < 3.0$.}
\end{figure}
\newpage
\begin{figure}
\vspace{-2cm}

\epsfysize=600pt \epsfbox{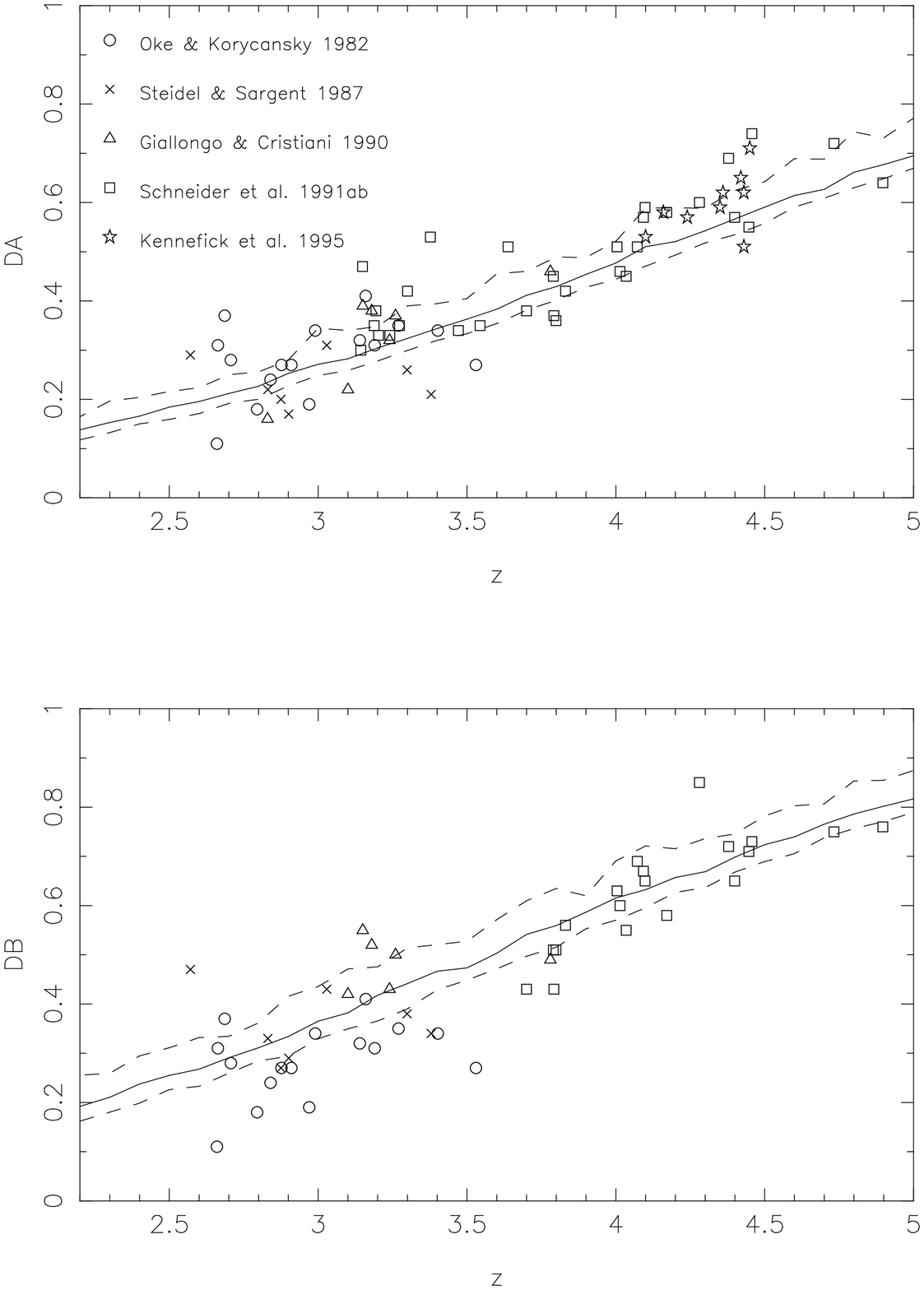}
\vspace{1cm}
Fig B1 -- Lyman depression $D_{A}$ and $D_{B}$ as a function of
redshift. Solid and dashed lines are the median and 68\% scatter from the simulation.

\end{figure}
\end{document}